\documentclass[twocolumn,12pt]{aa}

\usepackage{twoopt,natbib}
\usepackage{graphicx}
\usepackage[dvipsnames]{xcolor}
\usepackage{txfonts}
\usepackage[breaklinks=true,colorlinks=true,linkcolor=blue,citecolor=blue,filecolor=blue,urlcolor=blue]{hyperref}
\usepackage{multicol}
\usepackage{multirow}
\usepackage{amssymb}	
\usepackage{textcomp}
\usepackage{amsmath}
\usepackage{tabularx}
\usepackage[normalem]{ulem}
\usepackage{physics}
\usepackage{subcaption}
\usepackage{CJKutf8}

\begin{document}

   \title{Binary mass transfer in 3D - Mass transfer rate and morphology}
   \titlerunning{Roche-lobe overflow}

   \author{T. Ryu\inst{\ref{inst1},\ref{inst2},\ref{inst3}}, R. Sari\inst{\ref{inst4}}, S. E. de Mink\inst{\ref{inst1}}, O. David\inst{\ref{inst4}}, 
   R. Valli\inst{\ref{inst1}}, J.-Z. Ma \begin{CJK*}{UTF8}{gbsn}(马竟泽)\end{CJK*}\inst{\ref{inst1}}, S. Justham\inst{\ref{inst1}}, R. Pakmor\inst{\ref{inst1}}, \and H. Ritter\inst{\ref{inst1}}
          }
   \authorrunning{T. Ryu}
   \institute{Max-Planck-Institut f\"ur Astrophysik, Karl-Schwarzschild-Str. 1, Garching, 85748, Germany\label{inst1}
    \and JILA, University of Colorado and National Institute of Standards and Technology, 440 UCB, Boulder, CO 80308, USA\label{inst2}
    \and Department of Astrophysical and Planetary Sciences, 391 UCB, Boulder, CO 80309, USA\label{inst3}
    \and Racah Institute, Hebrew University of Jerusalem, Jerusalem, 91904, Israel\label{inst4}
                     }

   \date{Received XXX; accepted YYY}

 
  \abstract{
    Mass transfer is crucial in binary evolution, yet its theoretical treatment has long relied on analytic models whose key assumptions remain debated. We present a direct and systematic evaluation of these assumptions using high-resolution 3D hydrodynamical simulations including the Coriolis force. We simulate streams overflowing from both the inner and outer Lagrangian points, quantify mass transfer rates, and compare them with analytic solutions. We introduce scaling factors, including the overfilling factor, to render the problem dimensionless. 
    The donor-star models are simplified, with either an isentropic initial stratification and adiabatic evolution or an isothermal structure and evolution. However, the scalability of this formulation allows us to extend the results for a mass-transferring system to arbitrarily small overfilling factors for the adiabatic case. We find that the Coriolis force -- often neglected in analytic models -- strongly impacts the stream morphology: breaking axial symmetry, reducing the stream cross section, and shifting its origin toward the donor's trailing side. Contrary to common assumptions, the sonic surface is not flat and does not always intersect the Lagrangian point: instead, it is concave and shifted, particularly toward the accretor's trailing side. Despite these structural asymmetries, mass transfer rates are only mildly suppressed relative to analytic predictions and the deviation is remarkably small -- within a factor of two (ten) for the inner (outer) Lagrangian point over seven orders of magnitude in mass ratio. We use our results to extend the widely used mass-transfer rate prescriptions by \cite{Ritter1988} and \citet{KolbRitter1990}, for both the inner and outer Lagrangian points. These extensions can be readily adopted in stellar evolution codes like {\tt MESA}, with minimal changes where the original models are already in use.
  }
 

   \maketitle
%

\section{Introduction}
Many stars are born in binaries or even higher-order multiples \citep[e.g.,][]{Sana+2012,Sana+2013}. Members of sufficiently close binaries likely exchange mass with each other in their lifetime \citep{Paczynski1971} through the so-called Roche-lobe (RL) overflow, in which one binary member fills its RL and transfers mass to its companion. Mass transfer can significantly alter the evolution of both the accretor and donor \citep[e.g.,][]{Kippenhahn1969,Paczynski1971,NelsonEggleton2001}. In addition, mass exchange naturally leads to angular momentum exchange, affecting the binary orbit \citep[e.g.,][]{Huang1966,KippenhahnMeyer-Hofmeister1977} and the spin of the accretor \citep[e.g.,][]{deMink+2013}. Moreover, mass transfer can result in the generation of observable electromagnetic radiation as the transferred gas forms an accretion disk around the accretor \citep[e.g.,][]{Kruszewski1967,PrendergastBurbidge1968,King1996}. When mass transfer becomes unstable, the binary can undergo a common envelope phase, resulting in stellar mergers \citep[e.g.,][]{Ivanova+2013,deMink+2014,Schneider+2019,Ivanova+2020}. Consequently, mass transfer plays a direct or indirect role in the formation of many astrophysical phenomena and objects, such as X-ray binaries or millisecond pulsars \citep[e.g.,][]{JossRappaport1984,Tauris2006,Casares+2017,vandenHeuvel+2017}, blue stragglers \citep[e.g.,][]{Sandage1953,BurbidgeSandage1958,McCrea1964,Ferraro+2006,WangRyu2024}, cataclysmic variables \citep[e.g.,][]{Warner1995,Hellier2001}, type Ia supernovae \citep[e.g.,][]{WhelanIben1973,Nomoto1982,Han+2002,Rajamuthukumar+2024}, and gravitational-wave sources \citep[e.g.,][]{Belczynski+2002,Belczynski+2017,Marchant+2021,vanSon+2022,Tauris2023,DorozsmaiToonen2024}. 

Gas dynamics near a Lagrangian ($L$) point in RL overflow and the mass transfer rate has been studied mostly analytically. It is often assumed that the gas flow in stable mass transfer is steady, in which the Bernoulli theorem is satisfied along the stream lines in the corotating frame \citep{LubowShu1975}, neglecting the Coriolis force \citep[e.g.,][]{Paczynski+Sienkiewicz1972, MeyerMeyer-Hofmeister1983,Ritter1988,KolbRitter1990,PavlovskiiIvanova2015, HamersDosopoulou2019, Marchant+2021,Ivanova+2024}. In this case, gas near the donor's surface initially moves subsonically, reaches sonic speed near the $L$ point, and then accelerates to supersonic speeds beyond the $L$ point as it falls into the accretor's gravitational potential well \citep{LubowShu1975}.
These assumptions, along with the idea that the effective stream cross section is primarily determined by the shape of its density profile, enable the analytic computation of the overflow rate. While most existing analytic models are largely built upon this approach, there have been attempts to develop models that do not rely on the Bernoulli theorem, assuming a different stream morphology \citep[e.g.,][]{CehulaPejcha2023}. Some of these analytic models provide valuable insights into the mass transfer process and have been widely used in stellar evolution simulations \citep[e.g.][]{Marchant+2021}. However, some of their assumptions remain unconfirmed, and these models are limited in addressing potentially important physical effects, such as the Coriolis force.

As an alternative approach, hydrodynamical simulations have been used to study mass transfer, employing the smoothed particle hydrodynamics method \citep[e.g.,][]{ArmitageLivio1996, Lanzafame+2006,Norton+2008,Church+2009,LajoieSills2011,deVries+2014,ThomasWood2015,Bobrick+2017,Reichardt+2019} or Eulerian grid-based methods \citep[e.g.,][]{Bisikalo+1998,Oka+2002,DSouza+2006,ZhilkinBisikalo2010,Chen+2017,Kadam+2018,Toyouchi+2024,Dickson2024,Scherbak+2025}. These methods can easily incorporate multi-dimensional physical effects that analytic models do not account for. However, simulating stable mass transfer is computationally challenging, because resolving steep density gradients at the stellar surface and low-density overflowing gas requires prohibitively high resolution. In particular, to examine the stream morphology near the $L$ point, properly resolving both the stellar surface and overflowing streams is essential. Consequently, this resolution requirement sets a lower limit on the mass transfer rate or the overfilling factor that can be considered in 3D simulations. In most existing numerical work for RL overflow, the relative overfilling factor, defined as the relative fraction of the size of the donor to the effective size of RL, is $\gtrsim 0.1$, which is often high enough to lead to overflow through the outer $L$ point, depending on the binary parameters. Recently, \citet{Dickson2024} adopted a nonuniform set of paired grids to achieve remarkably high resolution, enabling simulations of RL overflow with a relative overfilling factor of 0.01 while properly resolving the overflowing streams. However, the cell size is still too large to reliably resolve the donor's surface. More recently, \citet{Scherbak+2025} performed 3D simulations of stable mass transfer, primarily aiming to investigate rapid mass transfer leading to circumbinary outflows. To this end, they considered high overfilling factors (1-10\%) and correspondingly high mass transfer rates. They adopted a modified spherical grid that concentrated resolution near the inner radial boundary and mid-plane; however, the resolution was not sufficiently high to fully resolve the donor's surface. 

While 3D simulations naturally capture non-linear hydrodynamical effects and the Coriolis force, they are limited to dynamical timescales and cannot easily explore a wide parameter space. Because mass transfer can persist over much longer evolutionary timescales, stellar evolution codes rely heavily on analytic prescriptions. For example, the most recent public version of the stellar evolution code {\tt MESA} \citep[r24.08.1;][]{paxton:13,Jermyn+2023} implements the original prescriptions by \citealt{Ritter1988} and \citealt{KolbRitter1990} \citep{Paxton+2015}, as well as an improved scheme for the \citet{Ritter1988} model ({\tt Arras} scheme) valid for a broader range of mass ratios. These methods estimate the overflow rate through the inner $L$ point, but they do not model outflow through the outer $L$ point. In systems with stable mass transfer in binaries with comparable masses, modeling overflow through the outer $L$ point is typically unnecessary. However, in systems undergoing unstable mass transfer -- especially those with rapidly expanding donors (for example, with convective envelopes) or extreme mass ratios -- accurately modeling outer $L$ outflow becomes essential, as it can occur frequently and significantly affect the binary evolution. This is particularly important given the potential observational outcomes of such systems, including stellar mergers \citep[e.g.,][]{Schneider+2019}, compact object binaries that are gravitational wave sources \citep[e.g.,][]{Belczynski+2002,Belczynski+2017}, and extreme mass ratio inspirals that emit both gravitational waves and electromagnetic radiation \citep[e.g.,][]{Olejak2025}.

In this work, we aim to gain a deeper understanding of the hydrodynamics of overflow through both the inner and outer $L$ points, using 3D hydrodynamics simulations with sufficient resolution to resolve the stellar surface and overflow streams. Instead of simulating the entire donor, we focus on the region near a $L$ point -- either the inner or outer point of the donor -- and solve the 3D hydrodynamics equations for that region in the corotating frame with the Coriolis force. We made simplified assumptions -- a polytropic relation for the envelope structure and an ideal-gas equation of state -- for analytic tractability and scalability, enabling a more direct comparison study with analytic solutions. We present simulation results for a wide range of mass ratios, defined as the donor-to-accretor mass ratio, ranging from $10^{-6}$ to $10$. By comparing the numerically estimated mass transfer rates to the analytic solutions of \citet{Ritter1988} and \citet{KolbRitter1990}, we provide correction factors to account for the differences between the two estimates, which can be readily implemented into stellar evolution codes like {\tt MESA}.

\begin{figure*}
    \centering
    \includegraphics[width=13cm]{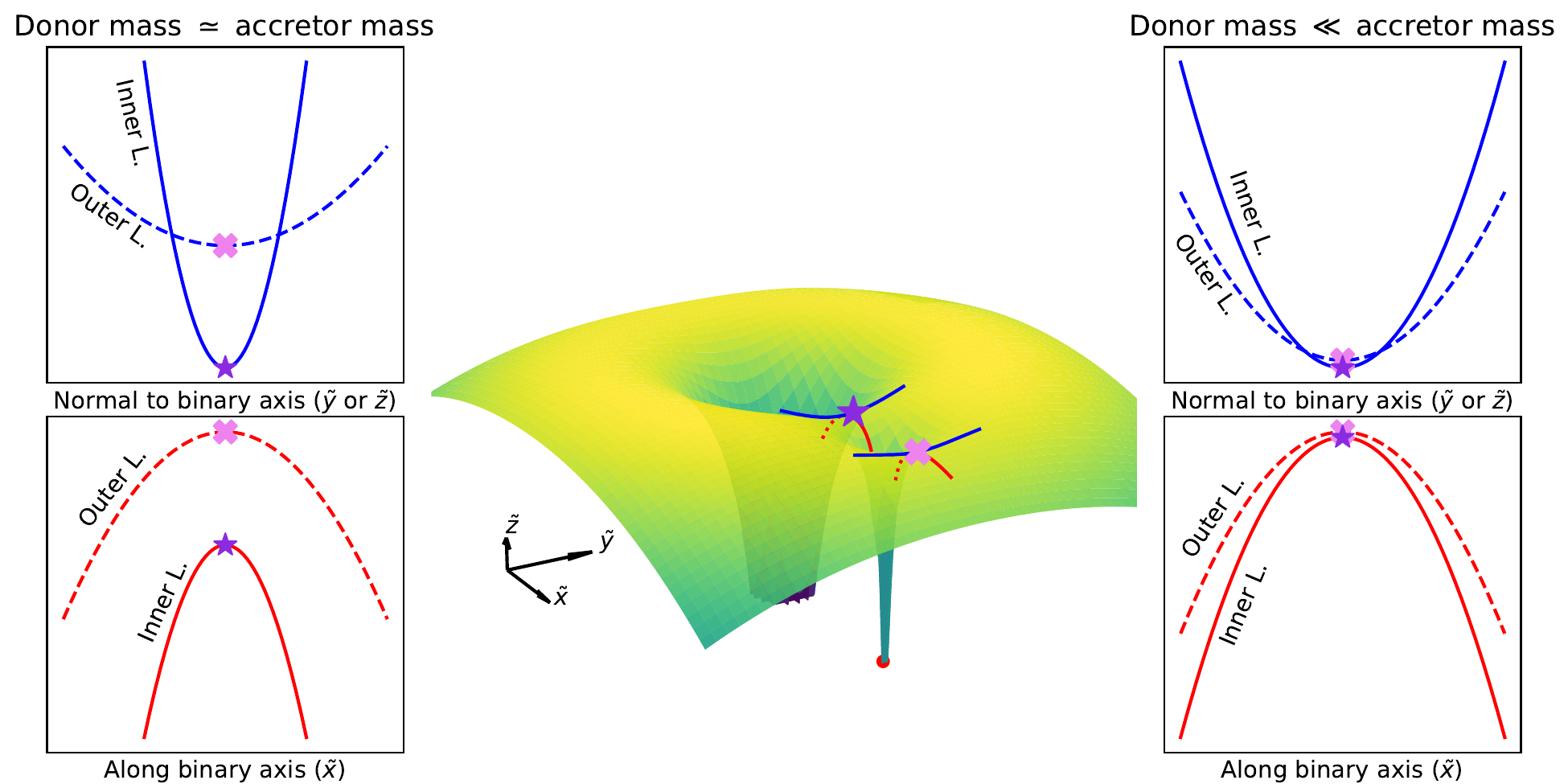}
    \caption{\label{fig:AB_schematic} Curvature of the gravitational (Roche) potential near the inner and outer Lagrangian points around the donor star. The middle panel depicts an overall shape of the Roche potential in a binary in the corotating frame. The inner and outer Lagrangian points of the donor, indicated by a violet 
 star and magenta cross, respectively, are where our computational domain is located. The $\tilde{x}$ axis of the domain is aligned with the binary axis and the $z$ axis is parallel to the binary orbit axis. The left panels illustrate the significantly steeper curvature and shallower potential depth -- both perpendicular (top) and parallel (bottom) to the binary axis -- near the inner Lagrangian point than the outer Lagrangian point for a system where the donor and accretor have comparable masses (for instance, stellar binaries). The right panels show the same quantities for a case where the donor mass is significantly smaller than the accretor mass (for example, stellar extreme mass ratio inspiral). In this case, similar to the comparable mass ratio case, the potential near the outer Lagrangian point exhibits shallower curvature. However, the depths become comparable. In fact, the depths become equal as the ratio of the donor mass to the accretor mass approaches zero. The curvature of the potential near the Lagrangian points is primarily determined by the coefficients of the quadratic terms in the Taylor expansion of the Roche potential (Eq.~\ref{eq:phi}).}
\end{figure*}

This paper is organized as follows. In Sect.~\ref{sec:problem}, we describe the Roche potential geometry near the inner and outer Lagrangian points, which becomes a crucial role in determining the rate difference between the two Lagrangian points. We then provide a summary of our main assumptions and our methodology in Sect.~\ref{sec:assumption}. In Sect.~\ref{sec:results}, we present our results for stream morphology (Sect.~\ref{subsec:morphology}), and mass transfer rates (Sect.~\ref{subsec:rate}). Based on our results, we extend the original analytic models by \cite{Ritter1988} and \cite{KolbRitter1990} to account for the potential geometry of both the inner and outer Lagrangian points, hydrodynamical effects, and the Coriolis force in Sect.~\ref{subsec:extension}. 
We discuss the astrophysical implications of our results in Sect.~\ref{sec:discussion}, a comparison of our simulation results to analytic models (Sect.~\ref{subsec:comparison}) and the stability of mass transfer (Sect.~\ref{subsec:stability}), and we conclude in Sect.~\ref{sec:conclusion}.

\section{Geometry of the Roche potential near the inner and outer Lagrangian points} \label{sec:problem}

Before discussing our methods and simulation results, we first examine the geometry of the gravitational potential around the Lagrangian point, which governs the motion of overflowing streams, and qualitatively discuss its implications for overflow. We begin with the classical approach \citep[e.g.,][]{Roche+1873, Kopal1959, Kruszewski1967, Eggleton2006}, in which the potential of a binary in its corotating frame -- assuming the binary members are point masses -- is given by
\begin{align}\label{eq:phi_exact}
    \phi_{\rm exact}(X,Y,Z)&=
    -{\frac{GM}{\left(X^2+Y^2 + Z^2\right)^{1/2}}}
    -{\frac{Gm}{\left((X-a)^2+Y^2 + Z^2\right)^{1/2}}}\nonumber\\
    &-\frac{1}{2} \Omega^{2}\left[ \left(X-{m \over M+m}a\right)^2+Y^2\right],
\end{align}
where $M$ and $m$ are the masses of the primary and secondary stars, respectively, $a$ is the semimajor axis of their relative circular orbits, $\Omega^2=G(M+m)/a^3$, $X$ is a coordinate along the line connecting the two masses, centered around $M$, $Y$ the perpendicular coordinate within the orbital plane, and $Z$ the coordinate along the orbital axis. 

Because we are interested in the potential geometry near the inner ($L_{\rm in}$) and outer Lagrangian ($L_{\rm out}$) points around a binary member (say the donor star), we introduce an approximate expression for $\phi$ by expanding it around the given $L$ point to second order. To do that, we define a new coordinate system ($x$, $y$, $z$) around a given $L$  point and expand $\phi_{\rm exact}$,
\begin{align}\label{eq:phi}
    \phi(x,y,z) &\simeq  \frac{1}{2}\left.\frac{\partial^{2} \phi_{\rm exact}}{\partial x^{2}}\right\rvert_{L}x^{2} + \frac{1}{2}\left.\frac{\partial^{2} \phi_{\rm exact}}{\partial y^{2}}\right\rvert_{L}y^{2} +\frac{1}{2}\left.\frac{\partial^{2} \phi_{\rm exact}}{\partial z^{2}}\right\rvert_{L}z^{2},\nonumber\\
    &= \frac{1}{2}\Omega^{2}(A x^{2} + B y^{2} + C z^{2}).
\end{align}
The first derivatives $\partial\phi_{\rm exact}/\partial_{x}|_{\rm L}=\partial\phi_{\rm exact}/\partial_{y}|_{\rm L}=\partial\phi_{\rm exact}/\partial_{z}|_{\rm L}$ are zero because the Lagrange point is a saddle point of the potential. Here, $A$, $B$, and $C$ are functions of mass ratio $q$ (see Fig.~\ref{fig:AB}) and $B = -(A+3)/2$ and $C=B+1= -(A+1)/2$ (see Equation 12 in \citealt{LubowShu1975}).

The geometry of the potential near the $L_{\rm in}$ and $L_{\rm out}$  points around the donor plays a major role in determining the morphology of overflowing streams and the overflow rate. This can be better understood from the coefficients $A$, $B$, and $C$ because they determine the curvature of the Roche potential, which is demonstrated in Fig.~\ref{fig:AB_schematic}. $A$ influences how easily gas near the donor's surface climbs the potential and how rapidly the gas accelerates after crossing the $L$ point (bottom panels). On the other hand, $B$ and $C$ determine the curvature of the potential in the direction perpendicular to the binary axis, intersecting the $L$ point (top panels). 

More specifically, for binaries with similar masses (for example, stellar binaries), the potential at the $L_{\rm in}$ point is much deeper and has a steeper curvature both parallel (larger $|A|$) and perpendicular (larger $B$ and $C$) to the binary axis than at the $L_{\rm out}$ point. This implies that the donor star must significantly overfill RL for $L_{\rm out}$ overflow to occur. For example, for equal-mass binaries, $L_{\rm out}$ overflow can occur when the size of the donor is larger than that of RL by $\simeq 30\%$ (or relative overfilling factor $\gtrsim 0.3$). On the other hand, for binaries with very different mass ratios (for instance, stellar extreme mass ratio inspirals), the depths of the potential at the $L_{\rm in}$ and $L_{\rm out}$ points become comparable, although the potential at the $L_{\rm in}$ point remains lower. This means $L_{\rm out}$ outflow can begin even when the donor slightly overfills the RL (for instance, relative overfilling factor $\gtrsim 10^{-3}$ for stellar extreme mass ratio inspiral with mass ratio of $10^{-6}$). However, the curvature of the potential near the $L_{\rm out}$ point is shallower, allowing for a larger cross-section of overflowing streams. This indicates that the $L_{\rm out}$ overflow rate can be comparable to or potentially larger than the $L_{\rm in}$ rate when the donor star overfills its RL sufficiently.

\begin{figure*}
    \centering
    \sidecaption
    \includegraphics[height=7.5cm]{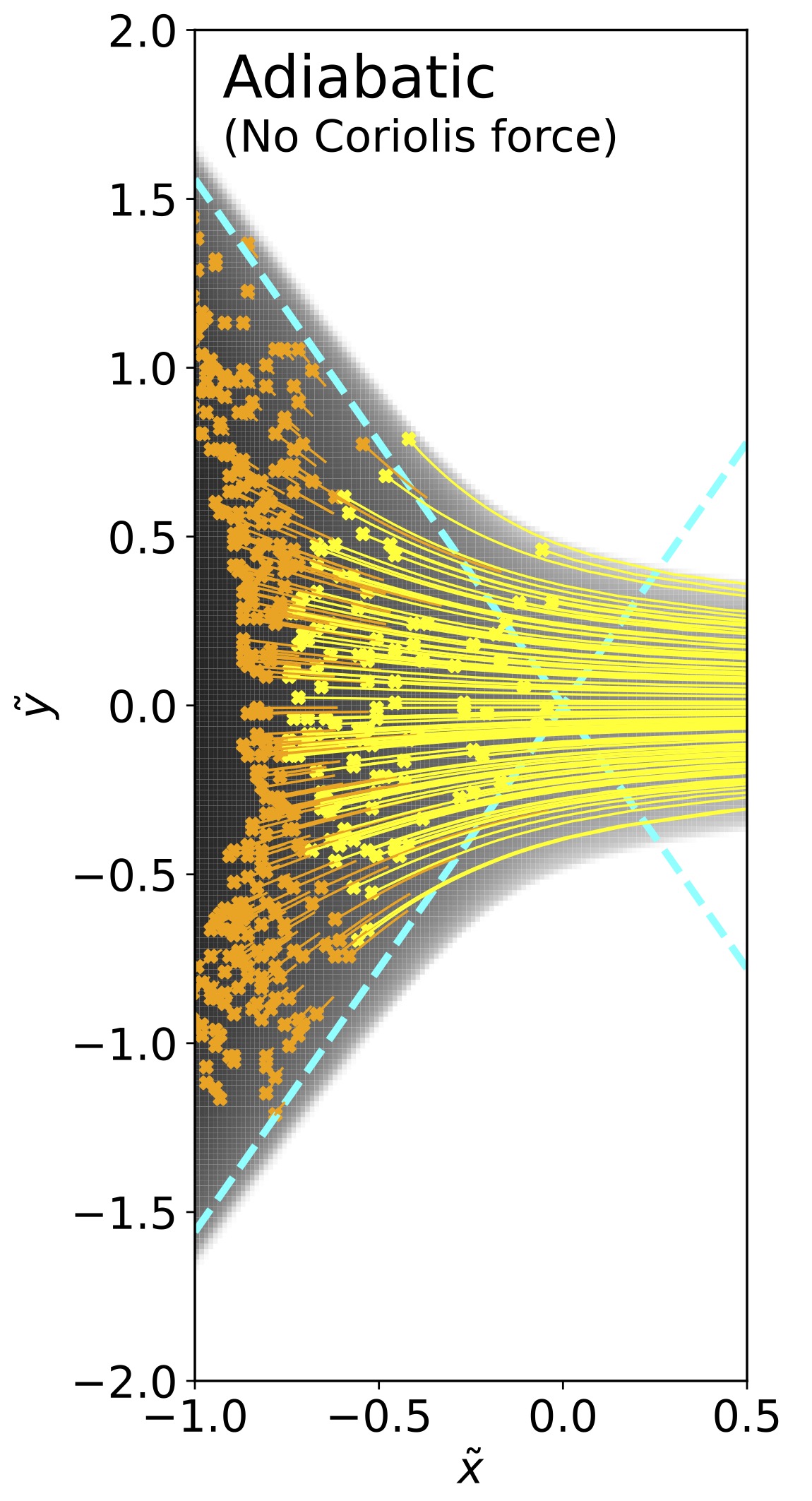}
    \includegraphics[height=7.5cm]{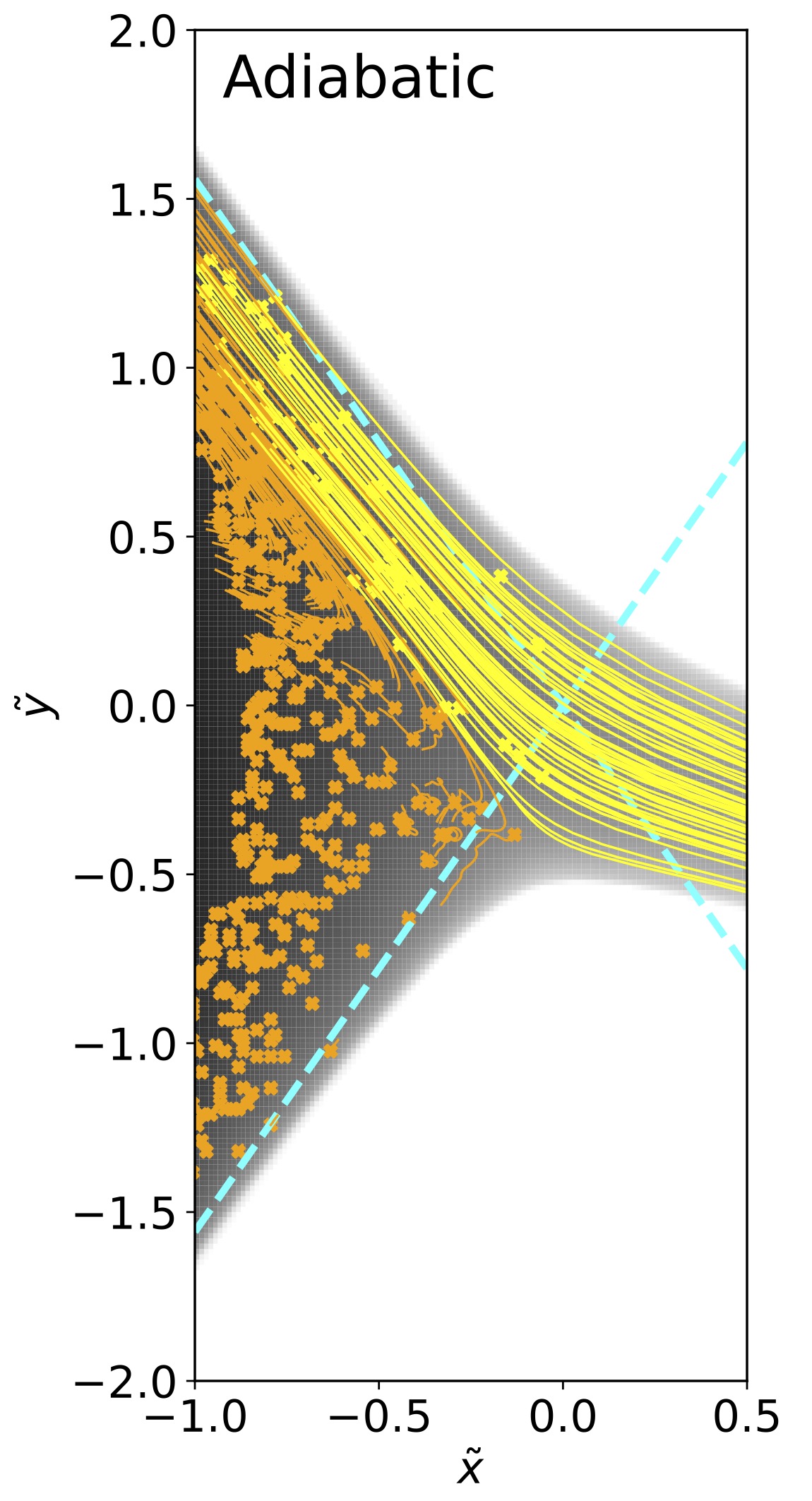}
    \includegraphics[height=7.5cm]{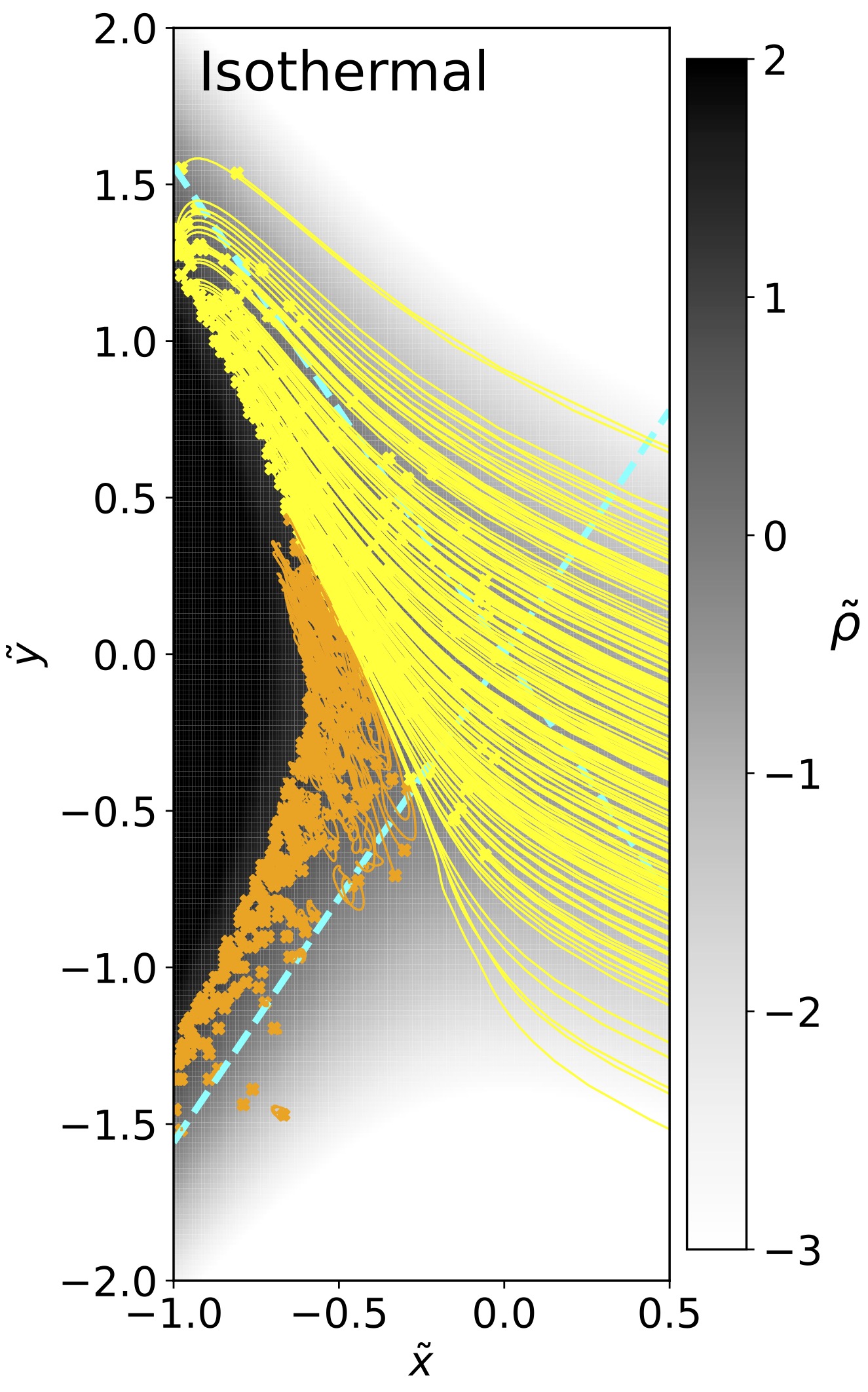}
    \caption{\label{fig:trajectory} Trajectories of particles for a steady state solution of an equal-mass adiabatic case (left: without Coriolis force and middle: with Coriolis force) and an equal-mass isothermal case (right) in the midplane around the inner Lagrangian point, plotted over the density distribution. The particles are initially distributed following the same distribution of the initial density profile over a region with $10^{-3}\lesssim\tilde{\rho}\lesssim 20$ and their trajectories are integrated for $\tilde{t}\simeq 20$ since the flow reaches a steady state, assuming the particles are advected with the gas. Crosses at one end of the lines indicate the initial locations of the particles. We distinguish particles that reach beyond the Lagrangian point (yellow) from those that remain inside the donor (orange) over the integration duration. }
\end{figure*}

\section{Summary of assumptions and numerical methods}\label{sec:assumption}

In this section, we provide a summary of key assumptions and numerical methods to perform 3D hydrodynamic simulations of mass transfer near the $L$ point, incorporating the approximate Roche potential -- {red} quadratic in position (Eq.~\ref{eq:phi}) -- and the Coriolis force in the corotating frame. We refer to Appendix Sect.~\ref{sec:analytic_solutions} and Sect.~\ref{sec:methods} for more details. 

We used the finite-volume adaptive mesh refinement magnetohydrodynamics code {\small ATHENA++} \citep{Stone+2008,Stone+2020}. Our Cartesian computational domain focuses on the region near each of the $L_{\rm in}$ and $L_{\rm out}$ points around the donor (see Fig.~\ref{fig:schematic} for a schematic diagram of the location and shape of the domain) The $x$ axis is aligned with the binary axis and the $z$ axis parallel to the orbit axis, with the coordinate center in the domain coinciding with the Lagrangian point. 

A key advantage of our methods is its scalability. We introduced scaling factors, including the RL overfilling factor, to make the hydrodynamics equations with the Roche potential and initial conditions dimensionless (see Sect.~\ref{subsec:scalingfactors} for the complete list of scaling factors). {red}In particular, the quadratic form of the potential allows defining a natural length scale for the problem as the square root of the potential difference. This means the problem retains the same dimensionless shape regardless of the exact value of the potential difference. In cases where the stellar surface is approximated to have a finite size (see the ``adiabatic'' case below), the potential difference corresponds to the overfilling factor, and the scalability enables generalizing simulation results for a given mass-transferring system to arbitrarily small overfilling factors. However, the accuracy of our results, when converted from code units to physical units, is influenced by the overfilling factor. Specifically, scaling to smaller overfilling factors yields more accurate results, as the approximated potential deviates more significantly from the exact potential at larger overfilling factors (see also Fig.~\ref{fig:codeunit_Rd}). Therefore, our approach is best suited for modeling the onset of mass transfer with small relative overfilling factors ($\lesssim 10^{-3}-10^{-2}$).

By focusing on overflow near the Lagrangian point with adaptive mesh refinement, we achieved unprecedented resolution: $20-80$ cells across the width of the stream crossing the Lagrangian point and $10-50$ cells per pressure scale height within the donor. This level of resolution for the overflow stream was achieved in previous simulations with relative overfilling factors greater than 0.01 \citep[e.g.,][]{Dickson2024}. However, the scalability of our method ensures that this high resolution is maintained even for arbitrarily small overfilling factors. In addition, to our knowledge, no previous simulations have achieved such a high resolution for both the donor's surface and the overflowing stream. We confirmed through resolution tests that the resolution is sufficiently high  to produce converged results (see Sect.~\ref{appendix:codesetup}).

We made two key assumptions: (1) the donor's surface before the onset of mass transfer is in hydrostatic equilibrium and follows a polytropic relation ($P\propto \rho^{\gamma}$ with a constant polytropic exponent $\gamma$, where $P$ is the pressure and $\rho$ is the density), and (2) the gas behaves as an ideal gas. While simplified, these assumptions are necessary for scalability and allow for fully analytical solutions for steady-state overflow (see Sect.~\ref{subsec:ana_mdot}) following \citet{Ritter1988} and \citet{KolbRitter1990}.  Their analytic models have been widely used in stellar evolution codes such as {\tt MESA}, but they account only for mass transfer through the $L_{\rm in}$ point. In this work, we extended their models to include mass transfer through the $L_{\rm out}$ point. By directly comparing analytic predictions with our simulation results for both the $L_{\rm in}$ and $L_{\rm out}$ points, we can achieve a critical assessment of the assumptions underlying analytic solutions, refine the mass transfer prescriptions for the $L_{\rm in}$ point, and provide prescriptions for the $L_{\rm out}$ point that can be readily implemented in stellar evolution codes.

 Under these assumptions, we first constructed an analytic solution for overcontact binaries in hydrostatic equilibrium (see Sect.~\ref{subsec:analytic}) and mapped it onto our numerical domain as the initial condition. We then applied an outflow boundary condition along a boundary facing the accretor (see Fig.~\ref{fig:schematic}). As a result, gas beyond the Lagrangian point -- toward the outflow-imposed boundary (namely, toward the accretor) -- quickly evacuates, while the donor on the opposite side begins transferring mass due to its initial RL overfilling.

We examine two cases:
\begin{itemize}
    \item Adiabatic: The initial structure follows a polytropic relation with $\gamma=5/3$, which evolves using an equation of state $P = (\Gamma-1)u$ with $\Gamma=5/3$. Here, $u$ is the internal energy density. Because of the same values of $\gamma$ (for the initial structure) and $\Gamma$ (for hydrodynamic evolution) and the absence of shocks, the entropy remains constant throughout the domain. This represents optically thick overflowing streams, such as the case with convective envelopes where the photosphere is outside the RL.
    \item Isothermal: The initial structure follows a polytropic relation with $\gamma=1$, which evolves assuming a constant $P/\rho$. This corresponds to the scenario where the transferring mass is optically thin (for example, photosphere inside the RL), maintaining the same sound speed. 
\end{itemize}
For each case, we simulated a wide range of mass ratio: $q=10^{-6}-10$ for the inner $L_{\rm in}$ and outer $L_{\rm out}$ Lagrangian points around the donor. Here, the mass ratio $q$ is defined as the donor-to-accretor mass ratio: For example, $q>1$ corresponds to overflow from a more massive donor to less massive accretor, while $q=10^{-6}$ corresponds to overflow onto a very massive accretor like a supermassive black hole. To examine the effect of the Coriolis force, we also performed an additional simulation for $q=1$ and $L_{\rm in}$ with $\gamma=5/3$, excluding the Coriolis force. We provide the complete list of our models in Table~\ref{tabl:models}.

\begin{figure}
    \centering
    \includegraphics[width=0.99\linewidth]{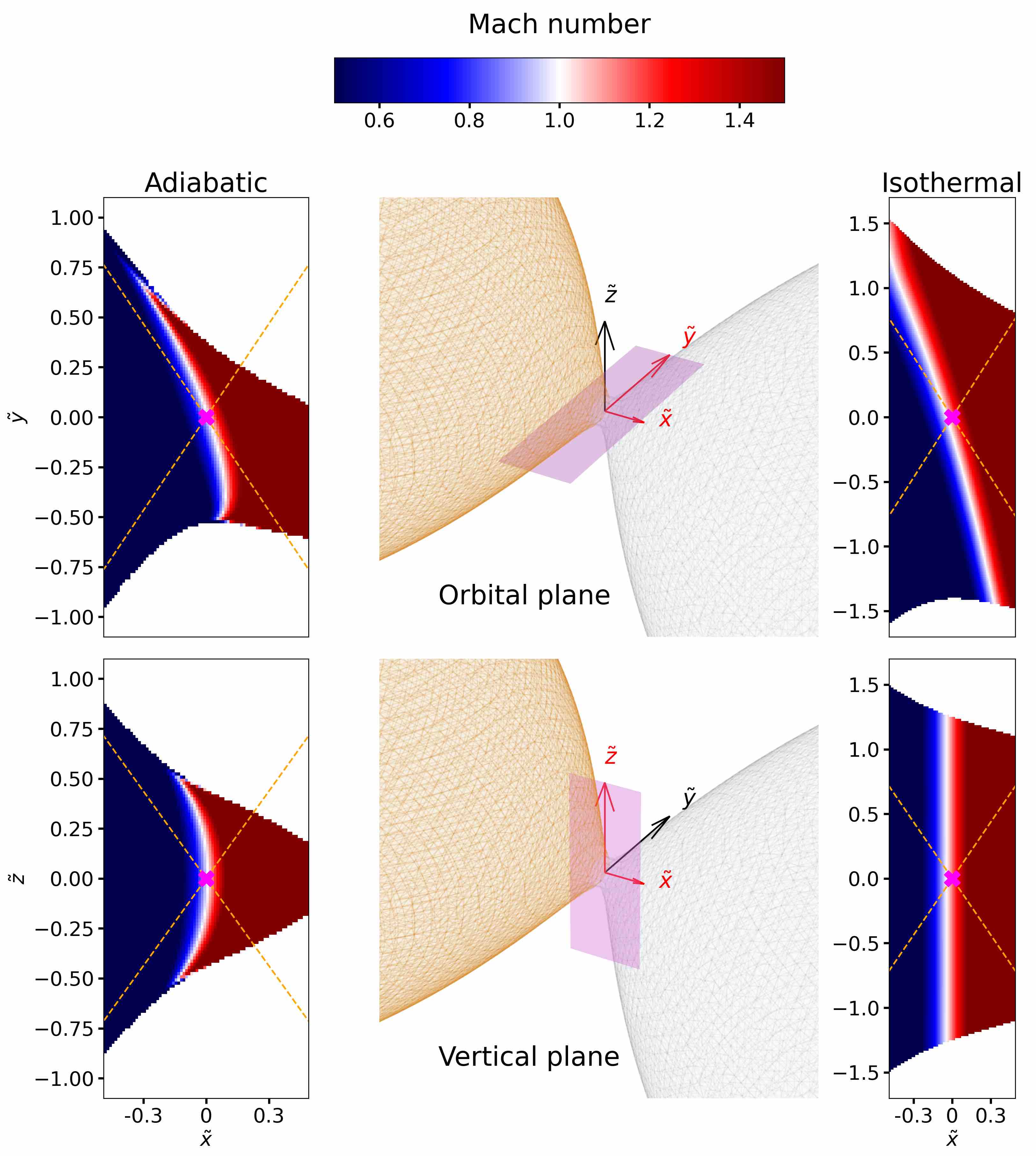}
    \caption{\label{fig:sonicpoint} Shape and location of the sonic surface near the inner Lagrangian point. The 3D images of the binary near the Lagrangian point in the center show the orientation of the 2D plane in which the 2D distribution of Mach number is depicted -- mid- (top) and vertical (bottom) slices. The panels on the left show the distributions for the equal-mass adiabatic case, while those on the right correspond to the equal-mass isothermal case. In the panels showing the Mach number distribution, the  dashed diagonal orange lines depict the shape of the Roche lobe. }
\end{figure}

\section{Results} \label{sec:results}
In this section, we present results for stream morphology, which we compare with assumptions made in analytic models, and for mass transfer rates, which we compare with the corresponding analytic predictions, whose full derivations are provided in Appendix Sect.~\ref{sec:analytic_solutions}. We provide results using dimensionless quantities scaled by characteristic scales of a given binary system, denoted by a tilde (for instance, $\tilde{t}$ and $\tilde{\rho}$). Conversion between scaled and physical units can be performed using the scaling factors given in Sect.~\ref{subsec:scalingfactors}. 

Turning on the outflow boundary condition leads to the sudden onset of a gas flow beyond the Lagrangian point. Due to the sudden change in state, the donor's surface and overflowing stream briefly undergo a transient phase. However, they soon reach a steady state, meaning no significant temporal evolution of gas morphology and the mass transfer rate. Therefore, we only focus on the morphology of streams and the mass overflow rate which have reached a steady state. For straightforward application of our results in stellar evolution and population synthesis codes, we provide fitting formulas for some of the main quantities using the identical mathematical form : $a + b \tanh[c\log_{10}(q) + d]^{e}$.

\subsection{Stream morphology}\label{subsec:morphology}

\begin{figure}
    \centering
    \includegraphics[width=0.99\linewidth]{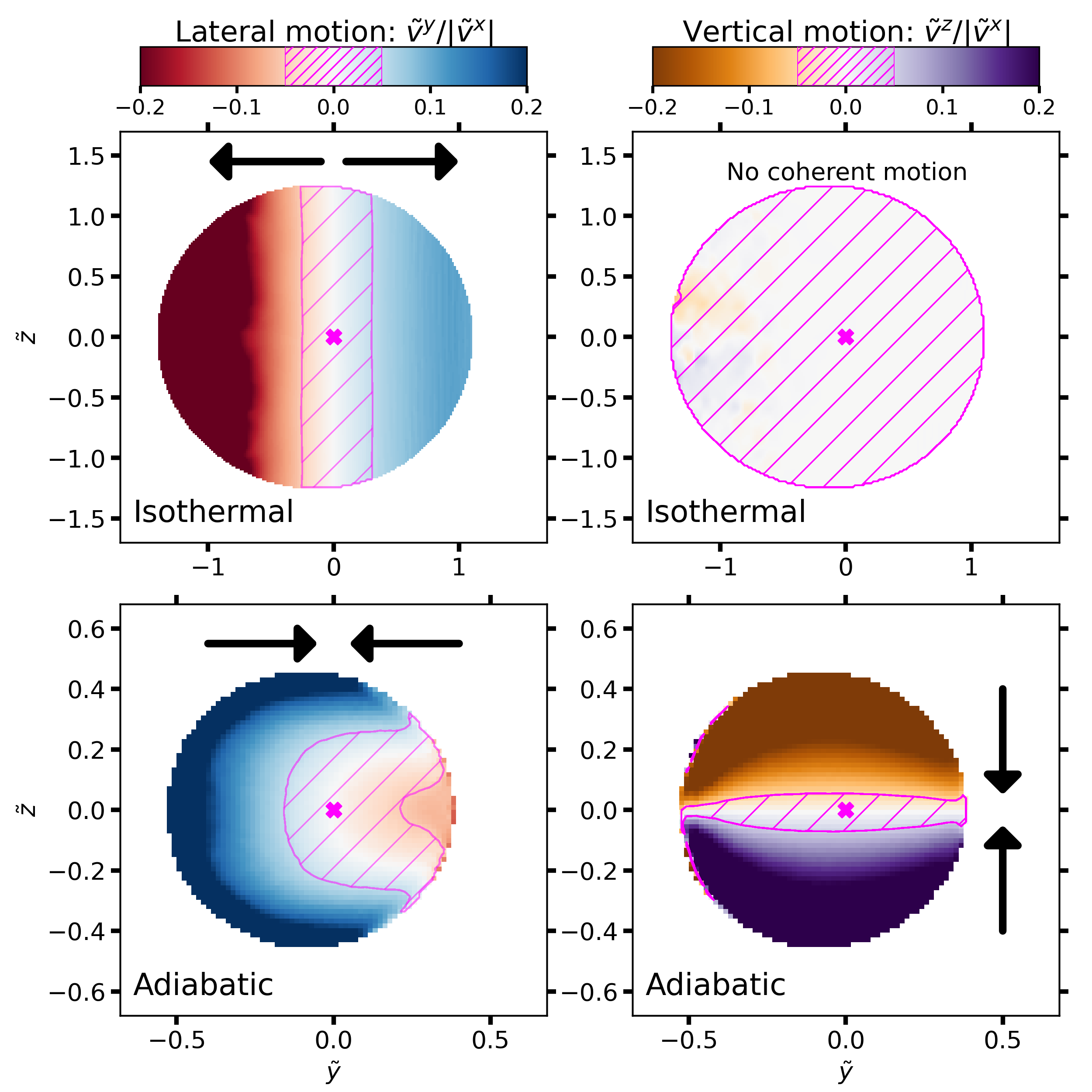}
    \caption{\label{fig:verticalmotion} Vertical and lateral motion of overflowing stream, relative to the motion toward the accretor, in the plane normal to the binary axis and intersecting the Lagrangian point (magenta cross), with the Coriolis force. Gas with $\tilde{\rho}<10^{-3}$ (white background), which contributes negligibly to the overflowing stream, is masked out for better visualization. $\tilde{v}^{i}$ ($i=x,y,z$) is the $i^{\rm th}$ velocity component relative to the Lagrangian point. In both cases -- adiabatic (left) and isothermal (right), we assume an equal-mass binary. The hatched areas indicate the velocity ratios less than 0.05, suggesting that the gas may be assumed to be in hydrostatic equilibrium. The arrows show the overall motion of the stream. }
\end{figure}

\subsubsection{Origin of the main stream}
In all our models including both adiabatic and isothermal cases, the main streams originate from the ``trailing'' side of the donor star along its orbital path because of the Coriolis force. We illustrate the origin of the main stream in Fig.~\ref{fig:trajectory} using the trajectories of tracer particles around $L_{\rm in}$ for $q=1$ in the midplane. The particles are initially distributed in a region of the donor with $10^{-3}\lesssim\tilde{\rho}\lesssim 20$, with their distribution scaling with the initial $\tilde{\rho}$. This ensures complete coverage of the region near the Lagrangian point and along the RL boundaries. Their trajectories are integrated for $\tilde{t}\simeq20$, after the system reached a steady state, under the assumption that the particles advect with the gas. 
It is worth noting that the deepest initial locations of the particles that have been transferred to the accretor largely depend on the integration duration. Consequently, this figure is intended to illustrate the overall shape of the main streams, rather than the maximum depth within the donor from which the gas can flow over the Lagrangian point. In the adiabatic case without the Coriolis force (left), the main stream originates from the middle. In contrast, in the same case with the Coriolis force (middle), the Coriolis force breaks the symmetry around the axis connecting the binary components (or ``binary axis'' at $\tilde{y}=0$), resulting in the main overflowing stream originating from the trailing side of the donor, and then being bent beyond the Lagrangian point toward the trailing side of the accretor. Similar features are found in the isothermal case with $q=1$ (right), and, indeed, in all our models. This main stream morphology resembles the stream motion near the Lagrangian point analytically predicted by \cite{LubowShu1975} (see their Fig. 3) and the streams observed in hydrodynamics simulations of the $L_{\rm in}$ streams for an equal-mass semidetached binary by \citet{Oka+2002}. 

\begin{figure*}
    \centering
    \includegraphics[width=0.48\linewidth]{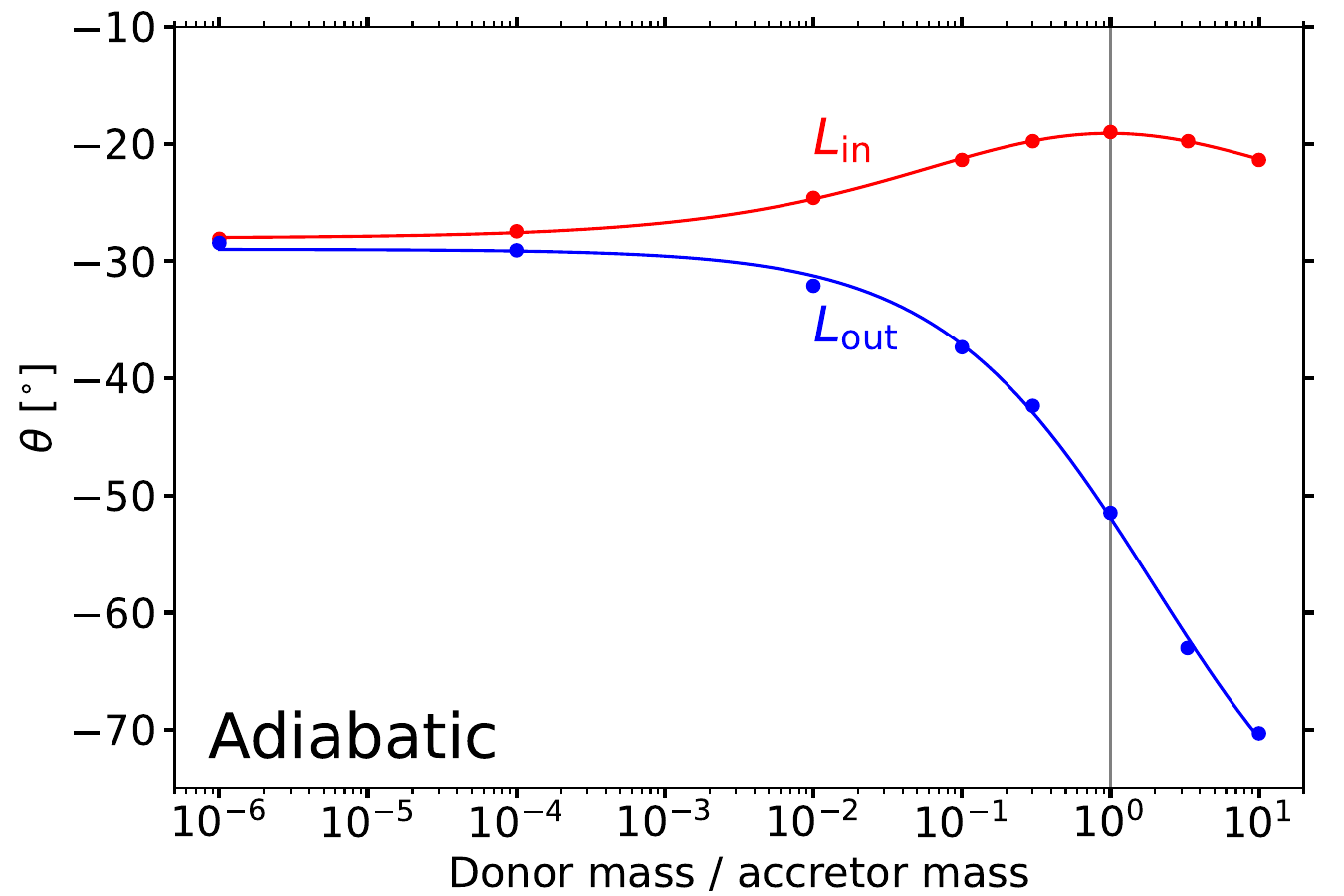}
    \includegraphics[width=0.48\linewidth]{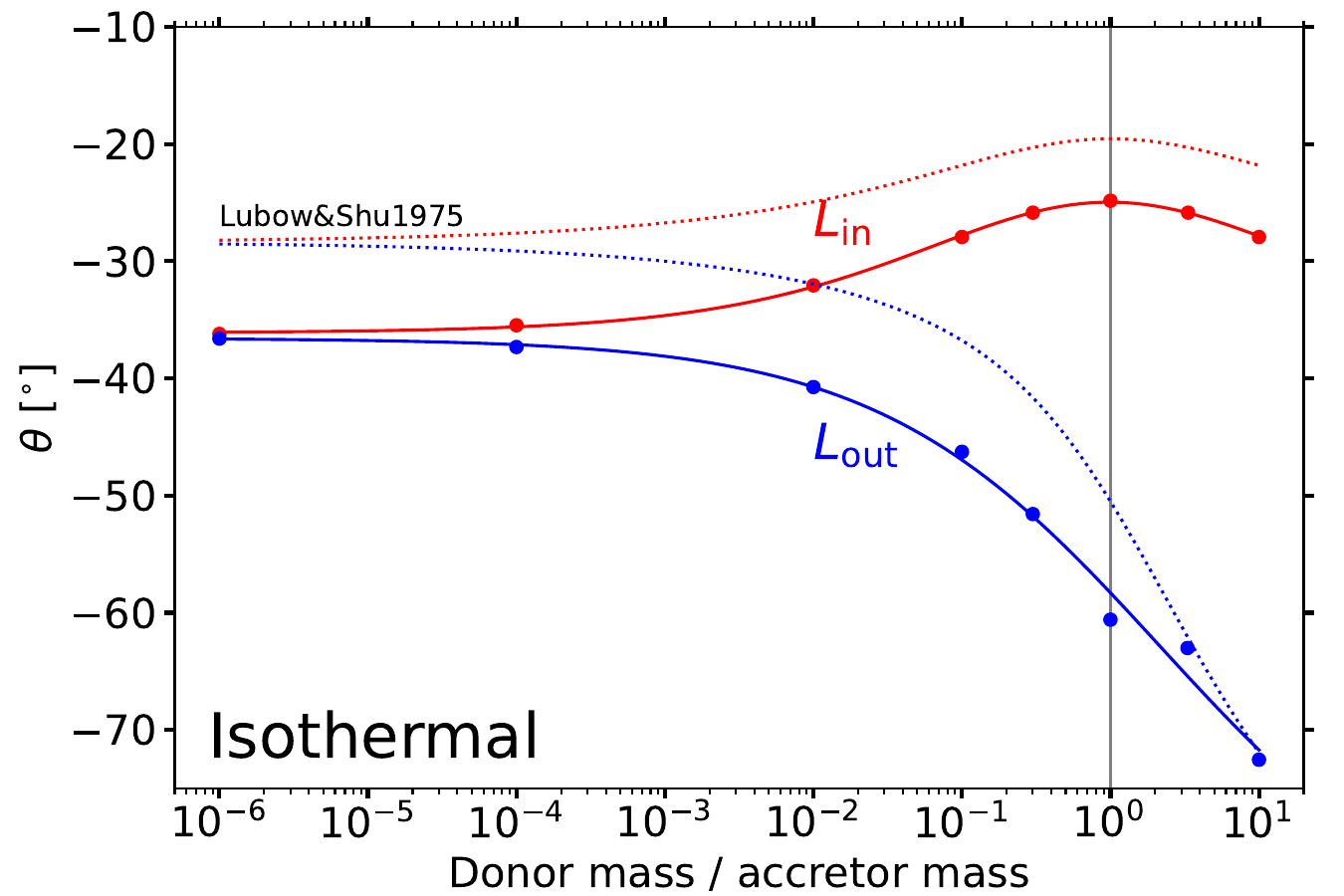}    
    \caption{\label{fig:theta} Angle of the transferring stream relative to the binary axis for adiabatic (left) and isothermal cases (right) in the presence of the Coriolis force. The angle is defined as $\tan\theta=\langle\tilde{v}^{y}\rangle/\langle\tilde{v}^{x}\rangle$, where $\langle\tilde{v}^{x}\rangle$ and $\langle\tilde{v}^{y}\rangle$ are mass-weighted averages of $\tilde{v}^{x}$ and $\tilde{v}^{y}$ measured at the right $x-$boundary. The solid curves following the dots indicate the fitting formulae given in Eq.~\ref{eq:fitting_theta}. The vertical solid gray lines indicate equal-mass binary. In the right panel, the dotted curves indicates the predictions of the tilt angle for the isothermal case by \citet{LubowShu1975}.  }
\end{figure*}

\subsubsection{Sonic surface}

The sonic surface -- the surface where the Mach number equals to unity -- is expected to be located in a very small neighborhood of the Lagrangian point \citep[for example, on the scale of $c_{\rm s}/\Omega$, where $c_{\rm s}$ is the sound speed at the surface of the donor;][]{LubowShu1975}. Analytic models \citep[e.g.,][]{Ritter1988, KolbRitter1990,Ge+2010,Jackson+2017,CehulaPejcha2023} or semi-analytic model \citep[e.g.,][]{PavlovskiiIvanova2015} typically estimate the mass transfer rate by assuming that the sonic speed is a flat surface perpendicular to the binary axis, intersecting the Lagrangian point. Therefore, the relative position of the sonic surface provides insight into how the assumptions about fluid properties in analytic models -- and, consequently, the mass flow rate -- differ from our numerical solutions.

We find that the sonic surface is not normal to the binary axis. In the adiabatic case, it is concave both vertically and laterally. Because of the Coriolis force, the sonic surface becomes asymmetric, stretching further toward the trailing side of the accretor, in the mid-plane, while remaining axisymmetric along the vertical direction. In the isothermal case, on the other hand, the sonic surface is vertically flat, but similar to the adiabatic case, it is laterally concave. This behavior is illustrated in Fig.~\ref{fig:sonicpoint}. The asymmetric, concave shape means that the fluid velocity along the plane normal to the binary axis is not the sonic speed, contrary to the assumption commonly made in analytic models. Even the sonic surface does not always intersect the Lagrangian point in both cases -- Mach number at the Lagrangian point $0.9 - 1$ for the adiabatic case and  1.1 - 1.3 for the isothermal case. This asymmetric shape of the sonic surface is also found in 3D hydrodynamics simulations of mass transfer in semi-detached binaries by \citet{Oka+2002} (see their Fig. 7) and \citet{Dickson2024} (see their Fig. 5). 

\subsubsection{Vertical and lateral motion of overflowing streams}
In analytic studies of isothermal mass transfer, it is often assumed that the overflowing stream at the inner Lagrangian point remains close to hydrostatic equilibrium along the vertical direction \citep{LubowShu1975} or along both vertical and lateral directions \citep{MeyerMeyer-Hofmeister1983,Ritter1988,CehulaPejcha2023}, allowing an analytic expression for the fluid density near the Lagrangian point. Furthermore, \citet{LubowShu1975} suggested the possibility of lateral expansion in the overflowing stream (see their Fig.3). In fact, we find that, in isothermal cases, the overflowing gas exhibits no significant vertical motion but does show lateral expansion, as assumed by \citet{LubowShu1975}. The expansion speed of most overflowing gas reaches up to a few tens of \% of the overflowing speed toward the donor (or $\tilde{v}^{v}/\tilde{v}^{x}\lesssim 0.2-0.3$). While the negligible vertical motion suggests that the gas is in hydrostatic equilibrium normal to the binary plane, the lateral expansion motion implies that the gas is not necessarily in hydrostatic equilibrium within the binary plane. This behavior is demonstrated in the top panels of Fig.~\ref{fig:verticalmotion} for the equal-mass, $L_{\rm in}$ mass transfer. On the other hand, in adiabatic cases, the main stream exhibits overall compressional motion toward the Lagrangian point, with speeds up to a few tens of \% of the overflowing speed toward the donor, which is illustrated in the bottom panels of Fig.~\ref{fig:verticalmotion}. The vertical and lateral motions in both cases can be understood qualitatively by considering how gas pressure responds to changes in density. As the gas overflows through the $L$ point, acceleration toward the accretor causes the overflowing stream density to decrease. In the isothermal case, because the temperature remains constant -- and thus the pressure decreases only linearly with density ($P\propto\rho$) -- the pressure can remain significant enough for the gas to expand or remain in hydrostatic equilibrium. In contrast, in the adiabatic case, the pressure drops more steeply ($P\propto\rho^{5/3}$), so the stream tends to contract due to the reduced thermal pressure. These trends are consistent across all models. 

\begin{figure*}
    \centering
    \includegraphics[width=0.49\linewidth]{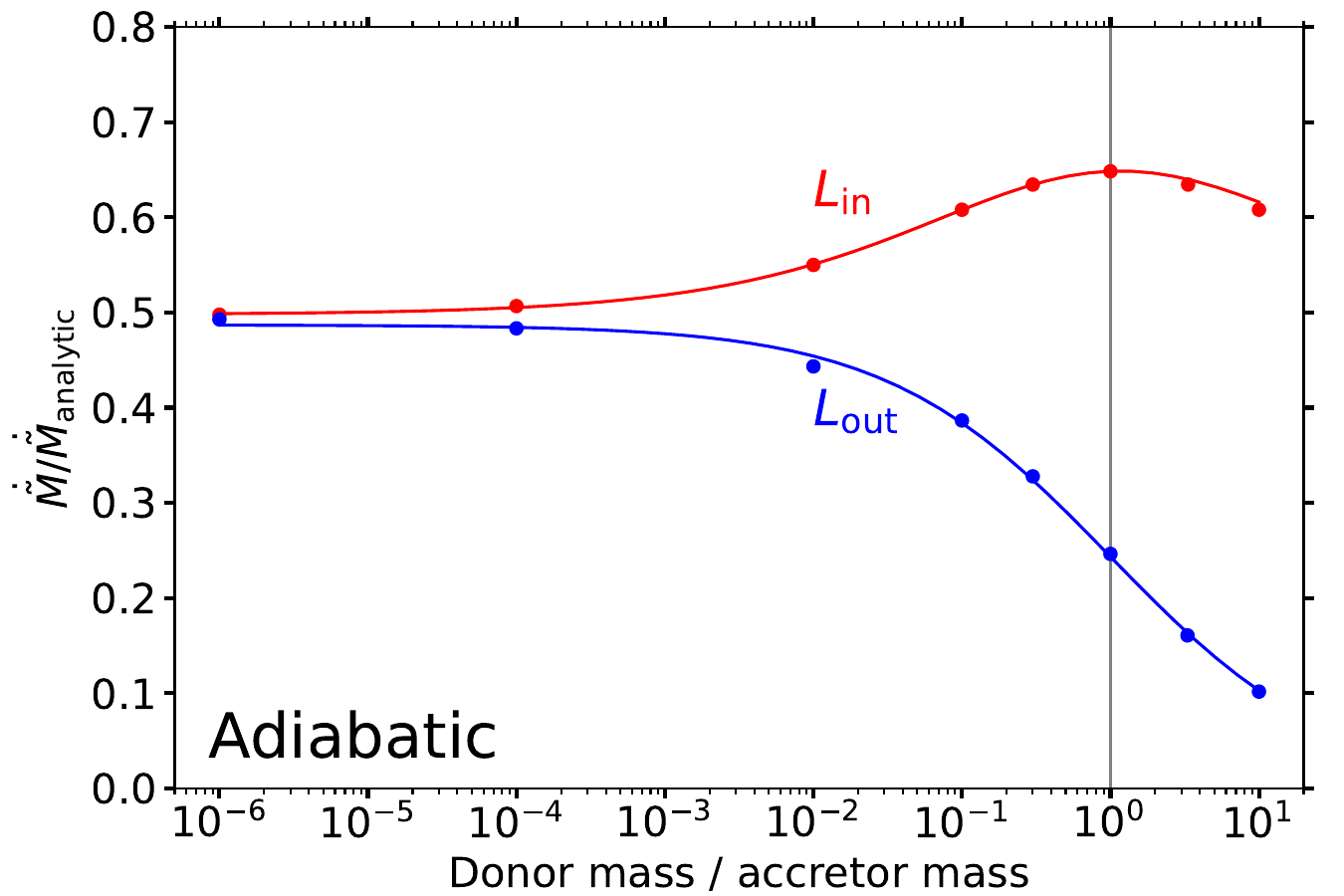}
    \includegraphics[width=0.49\linewidth]{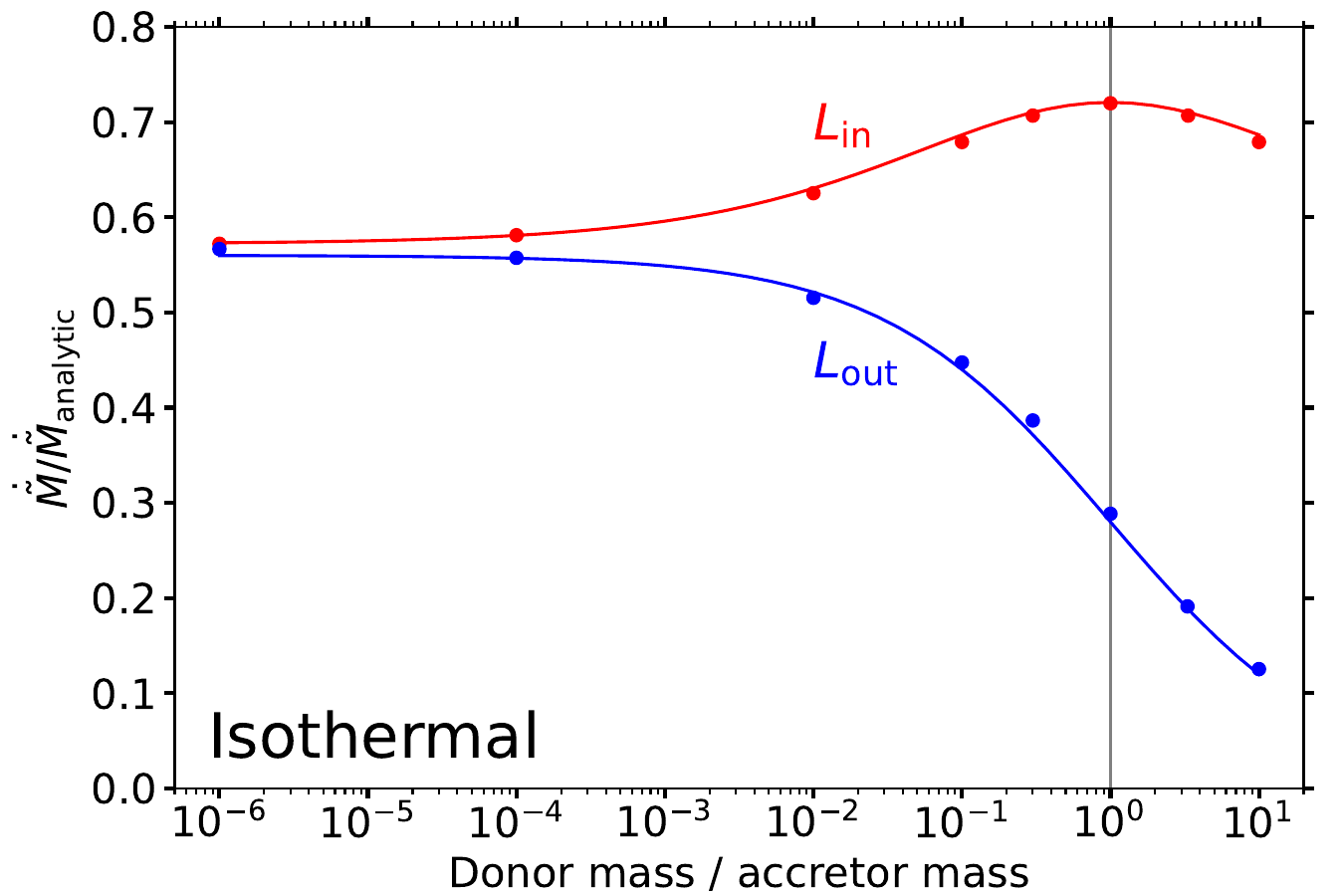}    
    \caption{\label{fig:mdot} Ratio of the numerically estimated mass transfer rate $\dot{\tilde{M}}$ in the presence of the Coriolis force to the analytic estimates $\dot{\tilde{M}}_{\rm analytic}$ (Eq.~\ref{eq:mdot_3Dadiabatic} for  for the adiabatic case (left) and Eq.~\ref{eq:mdot_3Disothermal}  for the isothermal case (right)). The solid curves following the dots indicate the fitting formulae given in Eq.~\ref{eq:fitting}. The solid vertical gray lines indicate equal-mass binary. Note that no assumption for the overfilling factor is made for this comparison as the overfilling factor is a scaling factor (see Eq.~\ref{eq:mdot_3Dadiabatic} and ~\ref{eq:mdot_3Disothermal}). We refer to Fig.~\ref{fig:mdotratio} for a comparison of $\dot{\tilde{M}}$ between $L_{\rm in}$ and $L_{\rm out}$ for a given overfilling factor.  }
\end{figure*}

\subsubsection{Tilt angle of overflowing gas}\label{subsub:tilt}
The Coriolis force bends the overflowing gas toward the trailing side of the accretor (see Fig.~\ref{fig:trajectory}). The tilt angle $\theta$ of the overflowing stream relative to the binary axis, defined as $\tan\theta=\langle\tilde{v}^{y}\rangle/\langle\tilde{v}^{x}\rangle$, is shown in Fig.~\ref{fig:theta} for the adiabatic (left) and isothermal (right) cases. Here, $\langle\tilde{v}^{x}\rangle$ and $\langle\tilde{v}^{y}\rangle$ is a mass-weighted average of $\tilde{v}^{x}$ $\tilde{v}^{y}$, respectively, across the right $x-$ boundary. For $L_{\rm in}$, $\theta$ varies within a relatively small range, peaking at $q=1$ and ranging from $-28^{\circ}$ to $-20^{\circ}$ in the adiabatic case and $-40^{\circ}$ to $-20^{\circ}$ for the isothermal case. On the other hand, for $L_{\rm out}$, $\theta$ decreases as $q$ increases, ranging from $-28^{\circ} (-36^{\circ})$ for $q=10^{-6}$ to $-72^{\circ} (-73^{\circ})$ for $q=10$ in the adiabatic (isothermal) case. 

This overall trend qualitatively follows the dependence of the RL intersection angle at the Lagrangian point relative to the binary axis, given by $\theta_{\rm RL} = \tan^{-1}(\sqrt{A/B})$. Here, $A$ and $B$ are the coefficients of the quadratic terms in $\tilde{x}$ and $\tilde{y}$, respectively, for the approximated Roche potential given in Eq.~\ref{eq:phi}. However, $\theta_{\rm RL}$ does not fully account for the trend quantitatively. The overall range of $\theta/\theta_{\rm RL}$ ranges from $0.6-0.3$ for $L_{\rm in}$ and $0.5-1$ for $L_{\rm out}$. In fact, \citet{LubowShu1975} estimated $\theta$ for the isothermal case using a perturbation analysis of the conservation equation for mass and momentum near the Lagrangian point, assuming vertical hydrostatic equilibrium. Their expression, $\cos2\theta=-4/3C+\sqrt{1-8/9C}$ (their Equation 24, dotted lines in Fig.~\ref{fig:theta})), consistently underestimates $\theta$ compared to our numerical estimates by $\lesssim20 - 30\%$ for all isothermal models\footnote{We find that the expression for the tilt angle from \citet{LubowShu1975} assuming isothermal gas reproduces the adiabatic simulation results much better -- within 3\%. However, the reason for this improved agreement is not clear.}.

We provide the fitting formula for $\theta$ in degrees with errors $\lesssim 4\%$,
\begin{align}\label{eq:fitting_theta}
    \theta= a + b[\tanh(c \log_{10}(q)+d)]^{e},
\end{align}
where for mass flow over the $L_{\rm in}$ point ($e=2$),
\begin{align}
    &a = -19.1,~ b = -8.92,~ c = 0.541,~d = 0\hspace{0.1in}\text{Adiabatic case},\nonumber\\
    &a = -25.0,~ b = -11.2,~ c = 0.558,~ d = 0 \hspace{0.1in}\text{Isothermal case},\nonumber
\end{align}
and for mass flow over the $L_{\rm out}$ point ($e=1$),
\begin{align}
    &a = -57.6,~ b = -28.6,~ c = 0.697,~ d = -0.202  \hspace{0.1in}\text{ Adiabatic case},\nonumber\\
    &a = -62.9, ~b =-26.3, ~c = 0.525, ~d = -0.176  \hspace{0.1in}\text{ Isothermal case}.\nonumber
\end{align}

\begin{figure*}
    \centering
    \sidecaption
    \includegraphics[width=0.6\linewidth]{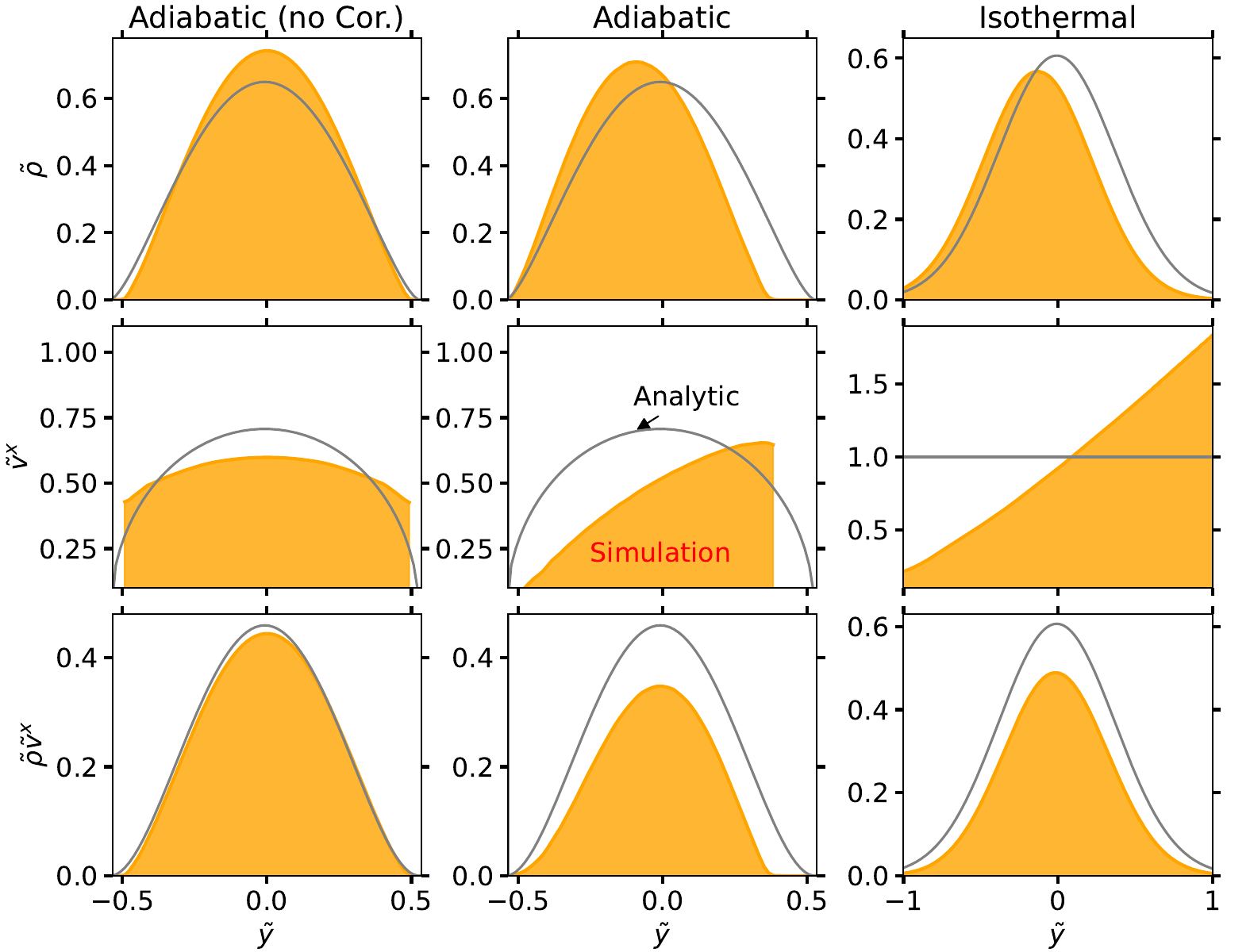}
    \caption{\label{fig:comparison} Comparison between numerical (orange) and analytic solutions (gray lines). We compare density $\tilde{\rho}$ (top row), velocity toward the accretor $\tilde{v}^{x}$ (middle row), and $\tilde{\rho}\tilde{v}^{x}$ (bottom row) from our simulations along the $\tilde{y}$ axis intersecting the inner Lagrangian point ($\tilde{y}=0$) to the corresponding analytic solutions. We consider equal-mass binaries for both adiabatic -- without (left column) and with (middle column) the Coriolis force -- and isothermal cases with the Coriolis force (right column).  }
\end{figure*}

\subsection{Mass transfer rate}\label{subsec:rate}

In Figure~\ref{fig:mdot}, we compare the mass transfer rate from our simulations to the analytic solution for the adiabatic (Eq.~\ref{eq:mdot_3Dadiabatic}) and isothermal cases (Eq.~\ref{eq:mdot_3Disothermal}). We estimate the mass transfer rate as $\dot{\tilde{M}}=\int\int\tilde{\rho}\tilde{v}^{x}d\tilde{y}d\tilde{z}$ at the right $x-$boundary. We find that the numerically computed mass transfer rates, which include the effects of the Coriolis force, are lower than the analytic estimates by $50-70\%$ for $L_{\rm in}$ and by up to $10 - 50\%$ for $L_{\rm out}$ across a wide range of $q$, from $10^{-6}$ to $10$. Notably, for the $L_{\rm in}$, equal-mass ($q=1$) adiabatic case, when the Coriolis force is excluded, the numerically estimated rate is only 4\% smaller than the analytic estimate. This suggests that the discrepancy between the numerical and analytic results is caused by the Coriolis force.

To understand the main cause of the discrepancy, we compare numerical solutions to analytic solutions for $\tilde{\rho}$, $\tilde{v}^{x}$, and $\tilde{\rho}\tilde{v}^{x}$ along the $\tilde{y}$ axis intersecting the $L_{\rm in}$ point ($\tilde{x}=0$) in the equal-mass cases separately to their analytic solutions in Fig.~\ref{fig:comparison}. Let us first examine the adiabatic cases without (left column) and with (middle column) the Coriolis force. Without the Coriolis force, the density profile is symmetric around $\tilde{y}=0$, as expected, but is generally larger than its analytic solution. Conversely, $\tilde{v}^{x}$ is smaller than its analytic solution by a similar amount. The higher $\tilde{\rho}$ and smaller $\tilde{v}^{x}$ almost perfectly cancel each other, resulting in $\tilde{\rho}\tilde{v}^{x}$ values that closely match the analytic solution.

When the Coriolis force is introduced (middle column of Fig.~\ref{fig:comparison}), the density profile shifts to negative $\tilde{y}$, while the stream boundary at $\tilde{y}=-0.5$ remains unchanged. This means that the stream is compressed -- the density peak is 1.1 times larger than that of the analytic solution -- and the overall stream width decreases, with the full width at half maximum being 1.2 times smaller than that of the analytic profile. Meanwhile, $\tilde{v}^{x}$ becomes asymmetric around $\tilde{y}=0$, increasing monotonically with $\tilde{y}$ within the stream and reaching  slightly above $0.6$. Consequently, $\tilde{v}^{x}$ is smaller than its analytic solution across more than 80\% of the stream's width. At the density peak ($\tilde{y}\simeq -0.1$), $\tilde{v}^{x}$ is smaller than its analytic solution by a factor of 1.5. Interestingly, despite the asymmetry in $\tilde{\rho}$ and $\tilde{v}^{x}$, their product, $\tilde{\rho}\tilde{v}^{x}$, exhibits a more symmetric profile, extending only slightly further toward negative $\tilde{y}$. Overall, $\tilde{\rho}\tilde{v}^{x}$ remains consistently smaller than its analytic counterpart. This suppression is primarily driven by the reduction in $\tilde{v}^{x}$ and effective stream cross section, evidenced by three key factors: 1) $\tilde{\rho}\tilde{v}^{x}$ remains suppressed despite the increase in $\tilde{\rho}$ for $\tilde{y}<0$, 2) on the opposite side, the suppression of $\tilde{\rho}$ only partially contributes to the reduction in $\tilde{\rho}\tilde{v}^{x}$, and 3) the asymmetry in $\tilde{\rho}$ and $\tilde{v}^{x}$ in the opposite sense effectively reduces the width of the stream. 

The density profile in the isothermal case, shown in the right column of Figure~\ref{fig:comparison}, is also shifted toward negative $\tilde{y}$ because of the Coriolis force, with its density peak lower than that of the analytic counterpart. For $\tilde{y}$ values smaller than the peak location, $\tilde{\rho}$ remains slightly above the analytic profile while, on the opposite side, $\tilde{\rho}$ is consistently lower. This suggests the stream is not compressed, but rather becomes less dense with a smaller width. The velocity profile is a monotonically increasing function of $\tilde{y}$ within the stream width ($-1\lesssim\tilde{y}\lesssim 1$). Similarly to the adiabatic case with the Coriolis force, despite the asymmetric shapes of $\tilde{\rho}$ and $\tilde{v}^{x}$, the product $\tilde{\rho}\tilde{v}^{x}$ remains symmetric around $\tilde{y}=0$ and is consistently smaller than its analytic counterpart. However, unlike the adiabatic case, it is difficult to determine which component primarily contributes to the suppression of $\dot{\tilde{M}}$ relative to $\dot{\tilde{M}}_{\rm analytic}$. The reduction in $\tilde{v}^{x}$ is the dominant factor for $\tilde{y}<0$, whereas for $\tilde{y}>0$, the suppression of $\tilde{\rho}$ plays a larger role.

The fact that the Coriolis force results in the suppression of $\dot{\tilde{M}}$ may suggest a possible correlation of the $\dot{\tilde{M}}$ suppression  with the overflow tilt angle ($\theta$), the RL intersection angle ($\theta_{\rm RL}$), or possibly both. The qualitative similarity in the dependence of these quantities on the mass ratio further strengthens the possibility of their connection. This is understandable, considering the Coriolis force essentially results in the main stream coming in parallel to the RL potential (so $\theta_{\rm RL}$) inside the donor before crossing the Lagrangian point. And beyond the Lagrangian point, the overflowing gas speed determines the strength of the Coriolis force exerted on it, tilting the overflowing stream and creating the non $\tilde{y}$-component of velocity (so $\theta$). 

It turns out that the correction by $\theta_{\rm RL}$, namely $\cos\theta_{\rm RL}$, reduces the discrepancy between the numerically and analytically estimated rates significantly more than the correction by $\theta$. Overall, the $\cos\theta_{\rm RL}$ correction (or $\cos\theta \dot{\tilde{M}}/\dot{\tilde{M}}_{\rm analytic}$) leads to smaller deviations from analytic solutions within up to a factor of 1.3 for $L_{\rm in}$ and $2$ for $L_{\rm out}$ in both the adiabatic and isothermal cases. When corrected with $\cos\theta$, the overall deviations from analytic solutions remain nearly the same as those without the correction because $\cos\theta$ is generally too small to make significant corrections. While the difference in the level of correction between the two angles is substantial, it is not entirely clear why $\theta_{\rm RL}$ serves as a better correction factor. 

Finally, we provide the fitting formula for $\dot{\tilde{M}}/\dot{\tilde{M}}_{\rm analytic}$ with errors $\lesssim 4\%$, sharing the identical functional form of the fitting formula for $\theta$ given in Eq.~\ref{eq:fitting_theta},
\begin{align}\label{eq:fitting_mdot}
    \frac{\dot{\tilde{M}}}{\dot{\tilde{M}}_{\rm analytic}}= a + b[\tanh(c \log_{10}(q)+d)]^{e},
\end{align}
where for mass flow over the $L_{\rm in}$ point ($e=2$),
\begin{align}
    &a = 0.649,~ b = -0.151,~ c = 0.540,~ d = -0.037\hspace{0.1in}\text{Adiabatic case},\nonumber\\
    &a = 0.721,~ b = -0.149,~ c = 0.522,~ d = 0,\hspace{0.1in}\text{Isothermal case},\nonumber
\end{align}
and for mass flow over the $L_{\rm out}$ point ($e=1$),
\begin{align}
    &a = 0.243, ~b = -0.243, ~c = -0.658, ~d = 0\hspace{0.1in}\text{Adiabatic case},\nonumber\\
    &a = 0.280, ~b = 0.280, ~c = -0.650, ~d = 0\hspace{0.1in}\text{Isothermal case}.\nonumber
\end{align}

\section{Proposed improvement for implementation in binary evolution codes}\label{subsec:extension}
In this section, we extend the analytic models by \citet{Ritter1988} and \citet{KolbRitter1990} to account for both the $L_{\rm in}$ and $L_{\rm out}$ points valid over a wide range of mass ratios. We provide a fitting formula (Eq.~\ref{eq:fitting_dynamicalterm}) below that can be readily implemented in stellar evolution codes such as {\tt MESA} where their original models are already in use. Outflow through $L_{\rm out}$ was considered by \citet{Marchant+2021}, who accounted for the proper geometry of the Roche potential near $L_{\rm out}$ based on the models of \citet{Ritter1988} and \citet{KolbRitter1990} \citep[see also ][]{Misra+2020}. However, we attempt to further incorporate the effects of hydrodynamics and the Coriolis force into their models. 

\citet{Ritter1988} and \citet{KolbRitter1990} developed analytic models for the $L_{\rm in}$ point with isothermal and adiabatic gas, respectively, assuming steady flow of  overflowing gas. In both cases, the mass overflow rate through the $L_{\rm in}$ point is given by 
\begin{align}\label{eq:analytic_mdot_generic}
    \dot{M}= \int_{\rm L} \rho_{\rm L} v_{\rm L} dA,
\end{align}
where $\rho_{\rm L}$ and $v_{\rm L}$ are density and speed at the Lagrangian point. The mass flux $ \rho_{\rm L} v_{\rm L}$ is integrated over the surface (the ``sonic'' surface) normal to the binary axis and intersecting the Lagrangian point, where the flow speed is assumed to equal the sound speed. The assumption of steady-state flow implies a constant Bernoulli constant (see Sect.~\ref{sec:analytic_solutions}), which analytically connects the density and sound speed at the photosphere, assumed to be in hydrostatic equilibrium, to those at the sonic surface. 

\begin{figure}
    \centering
    \includegraphics[width=9cm]{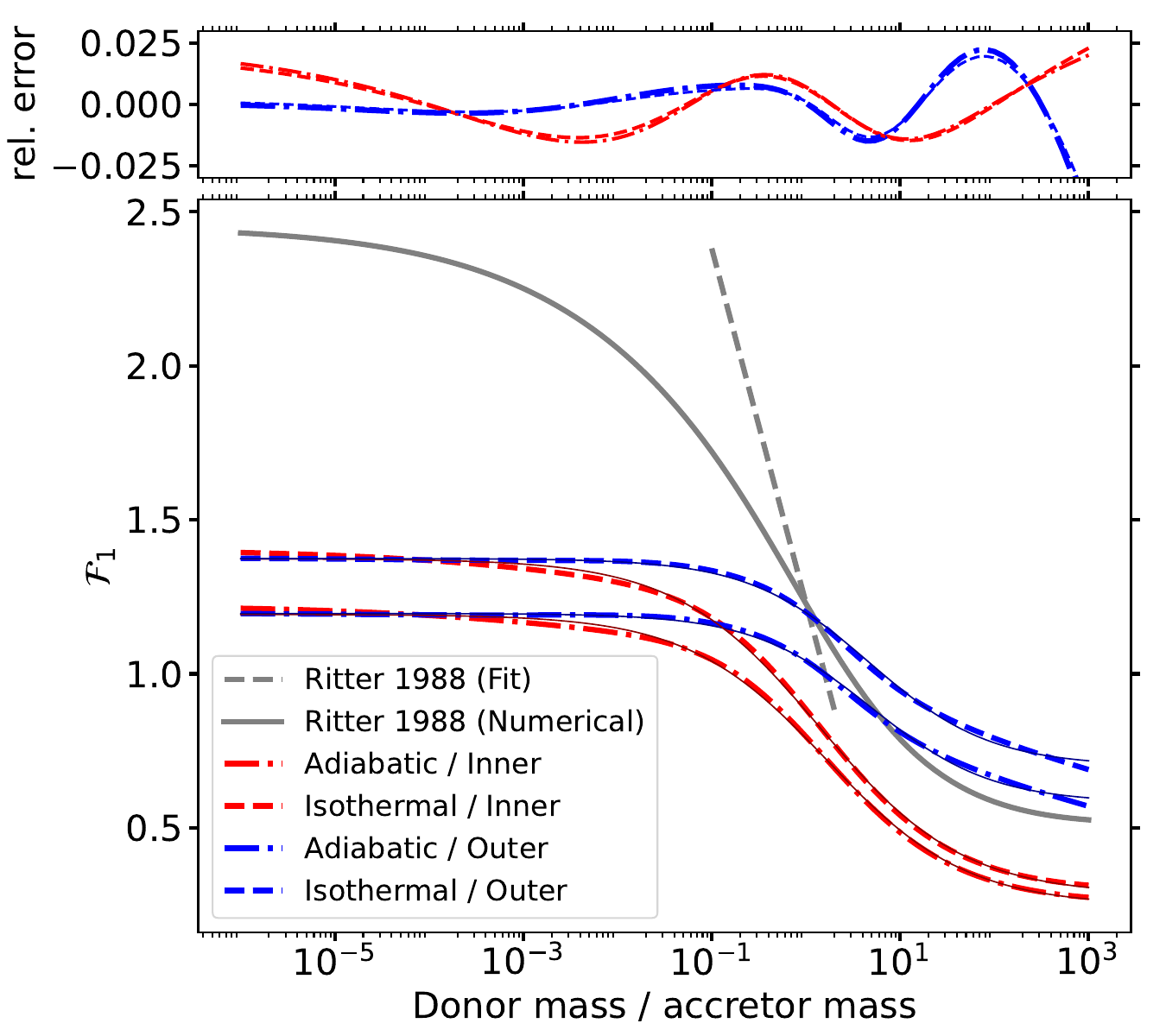}
    \caption{\label{fig:fitting}Dynamical term $\mathcal{F}_{1}$ accounting for the stream cross-section, hydrodynamical effects and the Coriolis force (colored lines) in the calculation of the mass transfer rate. In the bottom panel, the solid gray line represents the term only accounting for the potential geometry for the inner Lagrangian point introduced in \citet{Ritter1988}, who also provided a fitting formula (dashed gray line). The thick colored lines show the improved dynamical term for the adiabatic (dot-dashed lines) and isothermal (dashed lines) cases and the Roche potential geometry for the inner (red) and outer (blue) Lagrangian points. The fitting formula (thinner solid lines), closely overlapping with the thicker lines, reproduces the factors within 2.5\% across the full range of mass ratio considered.}
\end{figure}

In both the isothermal and adiabatic cases, the resulting integration contains the geometric term $F_{1}(q)$, following the notation of \citet{KolbRitter1990}\footnote{Also $F$ given in Eq. A8 in \citet{Ritter1988}.}, which accounts for the cross-section of the overflowing stream \citep[Eq. A2 for the isothermal case and A17 for the adiabatic case in][]{KolbRitter1990}. $F_{1}(q)$ depends on the effective size of RL and two coefficients of the quadratic terms in the Taylor expansion of the Roche potential. Specifically, $F_{1}(q)$ and two quadratic coefficients ($B$ and $C$ in Eq.~\ref{eq:phi_exact}) are related by
\begin{align}\label{eq:F1}
    F_{1}(q) = \frac{q}{1+q} \frac{f_{1}^{-3}}{\sqrt{BC}},
\end{align}
where $f_{1}$ is the effective size of RL in units of the semimajor axis.
Because no analytic expression for $F_{1}$ exists, \citet{Ritter1988} provided a fitting formula (their Eq. A9), valid for the inner Lagrangian point with mass ratios $0.1 \leq q\leq 2$. In Fig.~\ref{fig:fitting}, we depict the fitting formula (dashed gray line), which reproduces the numerically estimated  $F_{1}$ (solid gray line) within $40\%$ over the stated range of mass ratios. This fitting formula is used in the functions {\tt get\_info\_for\_ritter(), calculate\_kolb\_mdot\_thick()} to calculate the overflow rate in {\tt MESA} [version r24.08.1].

The easiest way to incorporate the correction factor for $\dot{M}$ discussed in Sect.~\ref{subsec:rate} may be to include it directly in $F_{1}$, using the ratio $\dot{\tilde{M}}/\dot{\tilde{M}}_{\rm analytic}$ from Eq.~\ref{eq:fitting}, so that 
\begin{align}
    \mathcal{F}_{1}(q) = \frac{q}{1+q} \frac{f_{1}^{-3}}{\sqrt{BC}}\frac{\dot{\tilde{M}}}{\dot{\tilde{M}}_{\rm analytic}}.
\end{align}
This new ``dynamical'' factor accounts not only for the different geometry of the $L_{\rm in}$ and $L_{\rm out}$ points (using $B$ and $C$ for each point), but also for the hydrodynamical effects and the Coriolis force, which is shown in Fig.~\ref{fig:fitting} as red ($L_{\rm in}$ point) and blue ($L_{\rm out}$ point) lines. We estimated $f_{1}$ for the equipotential of the
$L_{\rm in}$ point using Eq. 14 in \citet{Jackson+2017} and for that of the $L_{\rm out}$ point using Eq. 3 in \citet{Marchant+2021}. 

To further facilitate implementation, we also provide a fitting formula for $\mathcal{F}_{1}$ that only depends on the mass ratio, with an error $\lesssim 2.5\%$ in the range of $10^{-6}\leq q \leq 10^{3}$. The mathematical form of the fitting formula is the same for all four cases, and consistent with the form used in all previously provided fitting expressions:
\begin{align}\label{eq:fitting_dynamicalterm}
    \mathcal{F}_{1}(q)= a + b \tanh[c\log_{10}(q) + d],
\end{align}
where for mass flow over the $L_{\rm in}$ point,
\begin{align}
    &a = 0.720,~ b = 0.473,~ c = -0.676,~ d = 0.150\hspace{0.1in}\text{Adiabatic case},\nonumber\\
    &a = 0.827,~ b = 0.546,~ c = -0.659,~ d = 0.095\hspace{0.1in}\text{Isothermal case},\nonumber
\end{align}
and for mass flow over the $L_{\rm out}$ point,
\begin{align}
    &a = 0.889, ~b = 0.307, ~c = -0.786, ~d = 0.551\hspace{0.1in}\text{Adiabatic case},\nonumber\\
    &a = 1.037, ~b = 0.337, ~c = -0.778, ~d = 0.512\hspace{0.1in}\text{Isothermal case}.\nonumber
\end{align}

This implementation leads to three key advantages: (1) $\mathcal{F}_{1}(q)$ applies over a wider range of mass ratios; (2) $\mathcal{F}_{1}(q)$ accounts for hydrodynamical effects, Coriolis force, and the Roche potential geometry of both the $L_{\rm in}$ and $L_{\rm out}$ points in adiabatic and isothermal cases; and (3) in practice, implementing $\mathcal{F}_{1}(q)$ requires minimal revision to existing codes where $F_{1}$ is already in use. One only needs to replace the original expression (Eq.~\ref{eq:F1}) with the expression for $\mathcal{F}_{1}$ (Eq.~\ref{fig:fitting})\footnote{If the original models are implemented without including $F_{1}$, one can simply apply the correction factors given in Eq.~\ref{eq:fitting_mdot} to the final rate estimated from the original prescriptions.}. 

However, there are also some caveats to the extended prescriptions: (1) they are based on two simplified assumptions -- a polytopic relation for the donor's envelope and an ideal-gas equation of state. (2) Our approximation for the Roche potential becomes less accurate at larger overfilling factors. Although we cannot precisely determine the extent of the resulting error in the hydrodynamical evolution of overflowing streams and the overflowing rate, based on the relative errors in the potential values, caution is advised when applying our prescriptions to systems with overfilling factors greater than $10^{-3}-10^{-2}$. We plan to refine these prescriptions in a follow-up study using simulations that incorporate the full expression of the Roche potential and account for realistic stellar structure.

\begin{figure}
    \centering
    \includegraphics[width=0.95\linewidth]{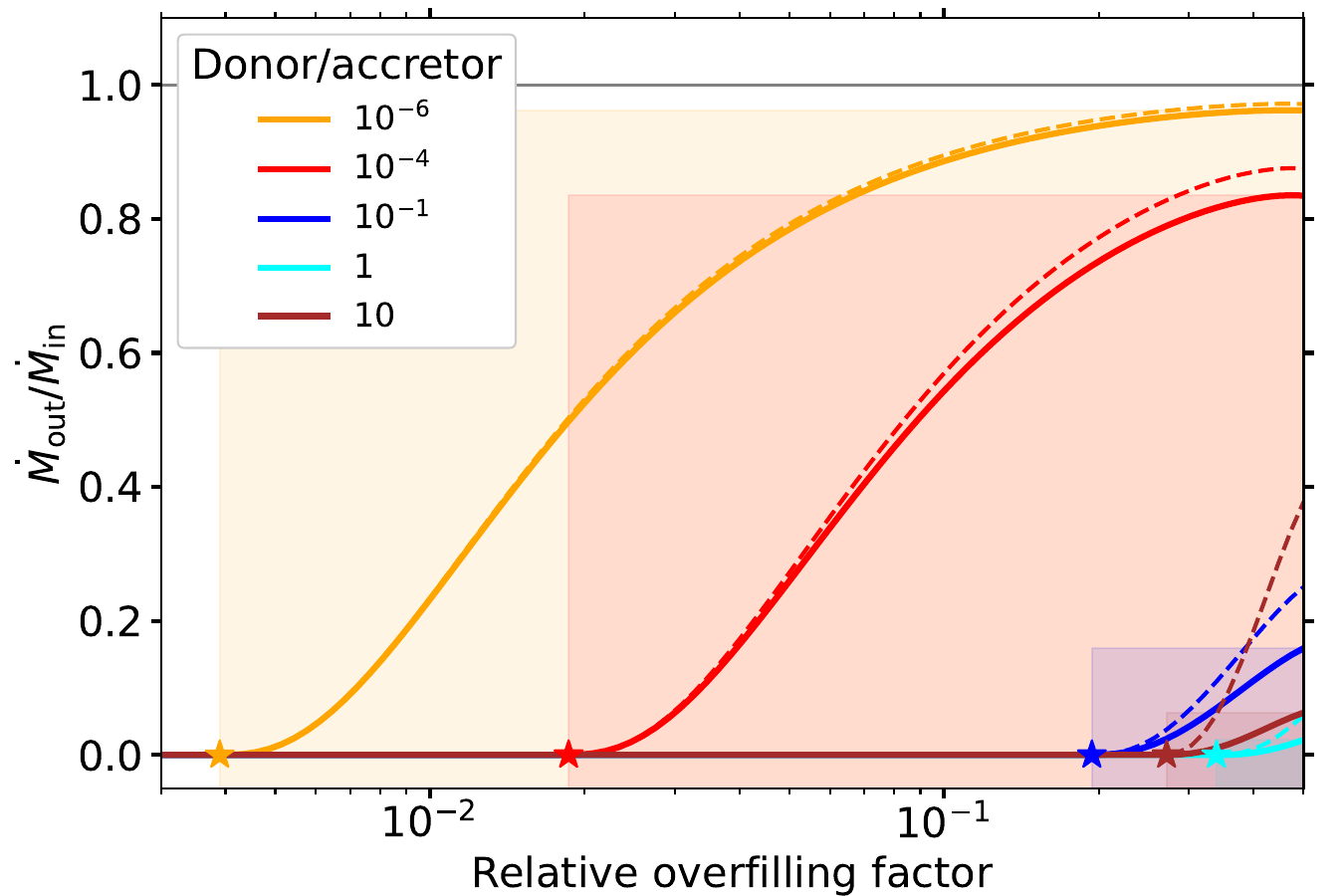}
    \caption{\label{fig:mdotratio} Onset of outflow through the outer Lagrangian point and the ratio of the mass transfer rate between the inner ($\dot{M}_{\rm in}$) and outer ($\dot{M}_{\rm out}$) Lagrangian points. The solid (dashed) lines represent the ratios predicted by the extension of \citet{KolbRitter1990} with (without) our correction factors. The star markers on the zero line indicate the characteristic relative overfilling factors at which outflow through the $L_{\rm out}$ point begins. The colored regions represent the range of overfilling factors for which outflow occurs through both Lagrangian points, as well as the maximum rate ratio over the range of parameters considered.   }
\end{figure}

Using these new prescriptions, we compare the $L_{\rm out}$ mass transfer rate to that for $L_{\rm in}$ in Fig.~\ref{fig:mdotratio}, as a function of the relative overfilling factor, assuming adiabatic gas\footnote{To compute the $L_{\rm out}$ and $L_{\rm in}$ rates at a given overfilling factor, we first determine the corresponding potential $\Phi_{*}$ using the volume-averaged potential given in \cite{Jackson+2017} (their Eq. 14). We then compute the potential differences from each Lagrangian point -- $\zeta_{\rm L,in} = \Phi_{*}-\Phi_{\rm L,in}$ and $\zeta_{\rm L,out}=\Phi_{*}-\Phi_{\rm L,out}$. If  $\zeta_{\rm L,out}\leq0$, no $L_{\rm out}$ outflow occurs so the $L_{\rm out}$ rate $\dot{M}_{\rm out}=0$. The characteristic overfilling factor at which $L_{\rm out}$ outflow begins (star markers in Fig.~\ref{fig:mdotratio}) corresponds to $\zeta_{\rm L,out}=0$. For $\zeta_{\rm L,out}>0$ when both $L_{\rm in}$ and $L_{\rm out}$ transfer occurs, we use the scaling $\dot{M}\propto |\zeta|^{3}$ (see Sect.~\ref{subsec:scalingfactors}) to compute $\dot{M}_{\rm out}/\dot{M}_{\rm in}$ as $(\dot{\tilde{M}}_{\rm out}/\dot{\tilde{M}}_{\rm in})|\zeta_{\rm L,out}/\zeta_{\rm L,in}|^{3}$, where $\dot{\tilde{M}}_{\rm out}$ and $\dot{\tilde{M}}_{\rm in}$ are the dimensionless rates for the outer and inner Lagrangian points, respectively, from our simulations.}. We also compare them to the ratios without our correction factor (dashed lines). The range of the relative overfilling factor considered in the plot is already worryingly high for our prescriptions to remain valid. However, we show this figure to demonstrate what our prescriptions can calculate. The characteristic overfilling factors for the onset of overflow through the $L_{\rm out}$ point are smaller for smaller mass ratios, as the potential depths at the $L_{\rm in}$ and $L_{\rm out}$ points become comparable (see Fig.~\ref{fig:AB_schematic})\footnote{These characteristic overfilling factors are determined solely by the relative depth of the potential at the Lagrangian points and are not related to our approximations for the shape of the Roche potential.}. Once $L_{\rm out}$ overflow begins, its rate increases more rapidly with overfilling than the $L_{\rm in}$ rate, leading to a rising trend with the overfilling factor. For $q=10^{-6}$, the $L_{\rm out}$  rate becomes comparable to the $L_{\rm in}$ rate at overfilling factors above 0.1 and our correction factor does not make a significant difference. We note that the rate ratio exceeds  unity when $q\lesssim 10^{-8}$, indicating the $L_{\rm out}$ outflow would dominate over the $L_{\rm in}$ mass transfer. For binaries with comparable masses, while the $L_{\rm out}$ overflow remains subdominant even at the overfilling factor of order unity, our correction factor can lead to a significant suppression of the $L_{\rm out}$ overflow relative to the $L_{\rm in}$ outflow. For example, for a mass ratio of 10, $\dot{M}_{\rm out}/\dot{M}_{\rm in}\lesssim 0.1$ with our correction factor, compared to $\dot{M}_{\rm out}/\dot{M}_{\rm in}\lesssim 0.4$ without it.

\begin{table*}
\caption{Comparison between analytic models and our simulation results}\label{tabl:comparison}
\centering
\small
\begin{tabular}{ c c | c c | c  c } 
\hline
\hline
                & & \multicolumn{2}{c|}{Isothermal} & \multicolumn{2}{c}{Adiabatic} \\
                  &                     & Analytic models     & our simulations                & Analytic models & our simulations \\
\hline
\multicolumn{2}{c|}{Steady-state overflow (\ref{com:steady})}  &  Yes &  Yes  & Yes & Yes\\
\hline
\multicolumn{2}{c|}{\multirow{2}{*}{Main stream morphology (\ref{com:origin})}} & \multirow{2}{*}{Axisymmetric} & Non-axisymmetric & \multirow{2}{*}{Axisymmetric} & Non-axisymmetric \\
                     &   & & (trailing side) & & (trailing side) \\
\hline
\multirow{2}{*}{Sonic surface (\ref{com:sonic})} & - Flat? & Yes & No - asymmetric concave & Yes & No - asymmetric concave\\
 &  - Intersecting $L$? & Yes & Not always & Yes & Not always\\
 \hline
\multicolumn{2}{c|}{Hydrostatic equilibrium (\ref{com:HSE})} &  \multirow{2}{*}{Yes$^{*}$} & Yes - normal to the orbital plane &  \multirow{2}{*}{-} &  No \\
\multicolumn{2}{c|}{normal to the binary axis}   &  & No - within the orbital plane &  & (converging toward $L$) \\
 \hline
 \hline
 \multicolumn{2}{c|}{Stream tile angle (\ref{com:angle})} & \multicolumn{2}{c|}{Agreement within 30\%}  & \multicolumn{2}{c}{-}\\
\hline
\end{tabular}
\tablefoot{ We compare five main assumptions (first four rows) and prediction (last row) made by analytic models to our simulation results for isothermal and adiabatic mass transfer. $L$ refers to the Lagrangian point. The references for each item are provided in the main text in Sect.~\ref{subsec:comparison}.\\
$^{*}$ \citet{LubowShu1975} argue that the stream near the $L$ point is in vertical hydrostatic equilibrium but undergoes lateral expansion, which agrees well with our simulation results.  }
\end{table*}

\section{Discussion} \label{sec:discussion}

\subsection{Comparison to previous analytic work}\label{subsec:comparison}
We compared our simulation results with predictions from prior analytic models in Sect.~\ref{sec:results}. For clarity, we summarize the key comparisons here and in Tab.~\ref{tabl:comparison}. The main assumptions of analytic models are:

\begin{enumerate}
    \item Steady-state overflow:\label{com:steady} Most models \citep[e.g.,][]{Paczynski+Sienkiewicz1972, LubowShu1975, Ritter1988, KolbRitter1990, Ge+2010,PavlovskiiIvanova2015, Jackson+2017, Marchant+2021} assume a steady-state overflow, enabling use of the Bernoulli theorem. Our simulations confirm this if the envelope lacks entropy stratification.
    
    \item Origin of the main stream:\label{com:origin} Analytic models using the Bernoulli theorem \citep[e.g.,][]{Paczynski+Sienkiewicz1972,Ritter1988,KolbRitter1990,Ge+2010,PavlovskiiIvanova2015, Jackson+2017,Marchant+2021} assume streams follow equipotentials from RL boundaries. An alternative view treats the $L$ point as a de Laval nozzle, with streamlines freely crossing equipotentials \citep{CehulaPejcha2023}. In this case, the main stream does not necessarily come from the sides, but rather from the middle along the binary axis. Our results resemble the nozzle picture without the Coriolis force, but with it, the stream originates mainly from the donor's trailing side.

    \item Sonic surface\label{com:sonic}: The Bernoulli theorem, along with the fact that the inner $L$ point is a saddle point, implies that the scale of the sound speed gradient should be comparable to the potential force gradient \citep{LubowShu1975}, leading to the conclusion that the fluid speed reaches the sonic speed at the $L$ point. In fact, this assumption also leads to the maximum steady-state flow through the Lagrangian point \citep{Paczynski+Sienkiewicz1972}. In later work \citep[e.g.,][]{MeyerMeyer-Hofmeister1983,Ritter1988,KolbRitter1990,Ge+2010,PavlovskiiIvanova2015,Jackson+2017,Marchant+2021}, the overflow rate is estimated assuming the flow speed in the flat plane normal to the binary axis and intersecting the Lagrangian point is the same as the local sound speed. Our simulations show that the sonic surface is not flat, but rather asymmetrically concave, which does not necessarily intersect the Lagrangian point.

    \item Hydrostatic equilibrium of the overflowing stream\label{com:HSE}: It is often assumed that the overflowing isothermal stream near the Lagrangian point remains close to hydrostatic equilibrium normal to the binary axis -- vertically \citep{LubowShu1975} or both vertically and laterally \citep{MeyerMeyer-Hofmeister1983, Ritter1988,Ge+2010,Jackson+2017,CehulaPejcha2023}. This assumption allows for the analytical derivation of a fluid density profile near the Lagrangian point (for instance, an exponential profile), which determines the streams' cross-section. We find that vertical (out-of-plane) hydrostatic balance holds, but in-plane equilibrium does not, which is consistent with \citet{LubowShu1975}. While this does not entirely invalidate the use of a hydrostatic density structure -- since the expansion speed reaches only up to tens of per cent of the overflowing speed -- it suggests that one should use caution when making such an assumption.

    \item Tilt angle of overflowing streams\label{com:angle}: The tilt angle of overflowing streams in the corotating frame is caused by the Coriolis force, which is often ignored in analytic models. \citet{LubowShu1975} estimated the title angle by investigating the asymptotic behavior of isothermal overflowing streams with the Coriolis force, which agrees with our simulation results to within 30\%.
\end{enumerate}

\subsection{Dynamical mass transfer stability}\label{subsec:stability}

Determining when mass transfer becomes unstable in binary evolution is critical to predicting the final outcomes of interacting binaries. One route to mass-transfer instability centers on how the donor star responds to mass loss during RL overflow, and whether the mass-transfer rate runs away to dynamical-timescale overflow. The canonical view was based on approximating the stellar structure with polytropes and assuming an adiabatic stellar response \citep[e.g.,][]{Paczynski1965,Hjellming+1987}.   
Those old adiabatic results remain widely used, despite the fact that calculations of mass transfer using detailed 1D stellar calculations find that thermal readjustment in the surface layers of the donor can help to significantly stabilize RL overflow \citep[e.g.,][]{Podsiadlowski+2002,WoodsIvanova2011,Temmink+2023}.

In the context of this work, the assumptions used in binary RL overflow calculations affect predictions for mass-transfer stability.  \citet{PavlovskiiIvanova2015}, \citet{Marchant+2021}, and \citet{CehulaPejcha2023} all adopted their respective updated schemes for treating Roche-lobe overflow and considered when the donor star became so large as to overflow $L_{\rm out}$, carrying away additional orbital angular momentum and potentially leading to unstable mass transfer (e.g.\, \citealt{Nariai+1976}; for other potential consequences of $L_{\rm out}$ overflow see also, for instance, \citealt{HubovaPejcha2019}).  The \citet{PavlovskiiIvanova2015} scheme tends to stabilize mass transfer, as investigated by \citet{Pavlovksii+2017}.  Conversely, \citet{CehulaPejcha2023} showed a massive-star example in which their mass-transfer scheme more easily led to $L_{\rm out}$ overflow than when using the schemes from \citet{KolbRitter1990} or \citet{Marchant+2021}.  Given the above, the results from our simulations relative to analytic estimates may be substantial enough to change the properties of final binary outcomes on an individual level. However, it remains unclear how the reduced mass transfer and enhanced angular momentum loss via $L_{\rm out}$ impact the statistical properties of a binary population, which requires a detailed population study.  

Our simulations indeed provide information on the instantaneous angular momentum carried by the transferred mass near $L_{\rm out}$. However, this would not be directly translated into true angular momentum loss. It is because the angular momentum of the transferred gas will continuously change due to the time-dependent potential of the binary until its motion becomes bound to an effectively single object, that is, either one of the binary components or the binary as a whole. Therefore, additional calculations are required to track the gas motion in order to quantify the true angular momentum loss. We will investigate this in a follow-up project.

\section{Conclusion and summary} \label{sec:conclusion}
We presented 3D hydrodynamical simulations of binary mass transfer in the corotating frame with the Coriolis force, spanning mass ratios from $10^{-6}$ to $10$ and considering both adiabatic and isothermal gases. The donor envelope followed a polytropic relation. Focusing on the vicinity of the inner and outer Lagrangian points, we achieved high resolution -- $10$–$50$ cells per scale height at the stellar surface and $20$–$80$ cells across the stream. Near the $L$ point, we used a second-order expansion of the Roche potential, allowing us to scale the problem with key parameters like the overfilling factor, and generalize results to arbitrarily small overfilling. 

Our main goals were (1) to study the stream’s morphology, tilt, and vertical/lateral structure, and (2) to estimate mass transfer rates and compare them with steady-state analytic models \citep[e.g.,][]{Ritter1988, KolbRitter1990}. The Coriolis force, which analytic models have often ignored, plays a central role in shaping the stream morphology, as well as determining the mass transfer rate. Our results can be summarized as:

$\bullet$ Properties of transferring streams: Without the Coriolis force, streams overflow symmetrically around the binary axis; with it, the main stream shifts to the donor’s trailing side, creating an asymmetric morphology. Accordingly, the sonic surface becomes asymmetrically concave, stretching further toward the trailing side of the accretor and not always intersecting the Lagrangian point. Tilt angles of the overflowing isothermal streams beyond the Lagrangian point agree with analytic predictions within 30\%. In adiabatic cases, gas compresses toward the Lagrangian point; in isothermal cases, the stream remains vertically hydrostatic but expands laterally. We compare these findings to analytic assumptions in Sect.~\ref{subsec:comparison}.

$\bullet$ Mass transfer rate: Simulated mass transfer rates with the Coriolis force  remains consistently smaller than analytic solutions for steady-state overflow through the inner Lagrangian point. However, the difference is remarkably small -- only by up to a factor of two across seven orders of magnitude in mass ratio. We extended the \citet{Ritter1988} and \citet{KolbRitter1990} models to include outer Lagrangian point overflow; these can be integrated into existing stellar evolution codes with minimal changes. Discrepancies for the outer Lagrangian point remain modest -- within a factor of ten for the same range of mass ratios. These differences primarily arise due to the Coriolis force, which reduces the stream cross-section and speed. Without the Coriolis force, simulations and analytic models agree within 5\%. The dependence of the rate on mass ratio aligns with the Roche-lobe intersection angle and stream tilt.

\vspace{0.1in}
Our 3D simulations provided numerical evidence to assess the validity of assumptions and predictions made in analytic models. In addition, we presented fitting formulas for key variables to improve existing models for the mass transfer rate and its stability. Although the results from our simulations cannot directly address the long-term effects of the corrected mass transfer rate on binary evolution, the corrections may significantly enhance the reliability of mass transfer models, including stability criteria. We employed polytropic relations for the stellar structure primarily for analytic tractability and comparison with analytic solutions. However, this assumption, along with the use of a simple equation of state, does not fully capture the complexity of realistic stellar models. In future work, we will improve our models by incorporating more realistic conditions, such as realistic envelope structure, an equation of state that accounts for partial ionization, convective motion, radiative processes, and magnetic fields. 

\begin{acknowledgements}
TR is grateful to Norbert Langer and Ond\v{r}ej Pejcha for useful discussions. TR thanks Alejandro Vigna-G\'omez and Jakub Klencki for useful comments on the manuscript.The authors thank the anonymous referee for their constructive feedback and suggestions. The authors acknowledge support from the German Israel Foundation for Scientific Research \& Development Research Grant No. I-873-303.5-2024. RS is supported by an ISF, BSF/NSF and MOST grants, and a Humboldt Fellowship. This research was supported in part by grant NSF PHY-2309135 to the Kavli Institute for Theoretical Physics (KITP). The simulations presented in this paper were conducted using computational resources (and/or scientific computing services) at the Max-Planck Computing \& Data Facility. The authors gratefully acknowledge the scientific support and HPC resources provided by the Erlangen National High Performance Computing Center (NHR@FAU) of the Friedrich-Alexander-Universität Erlangen-Nürnberg (FAU) under the NHR projects b166ea10 and b222dd10. NHR funding is provided by federal and Bavarian state authorities. NHR@FAU hardware is partially funded by the German Research Foundation (DFG) – 440719683. In addition, some of the simulations were performed on the national supercomputer Hawk at the High Performance Computing Center Stuttgart (HLRS) under the grant number 44232.
\end{acknowledgements}

%
  \bibliographystyle{aa} 
  \bibliography{RLOF} 
%

\begin{appendix}

\section{Analytic solution} \label{sec:analytic_solutions}

\begin{figure*}
    \centering
    \includegraphics[width=0.48\linewidth]{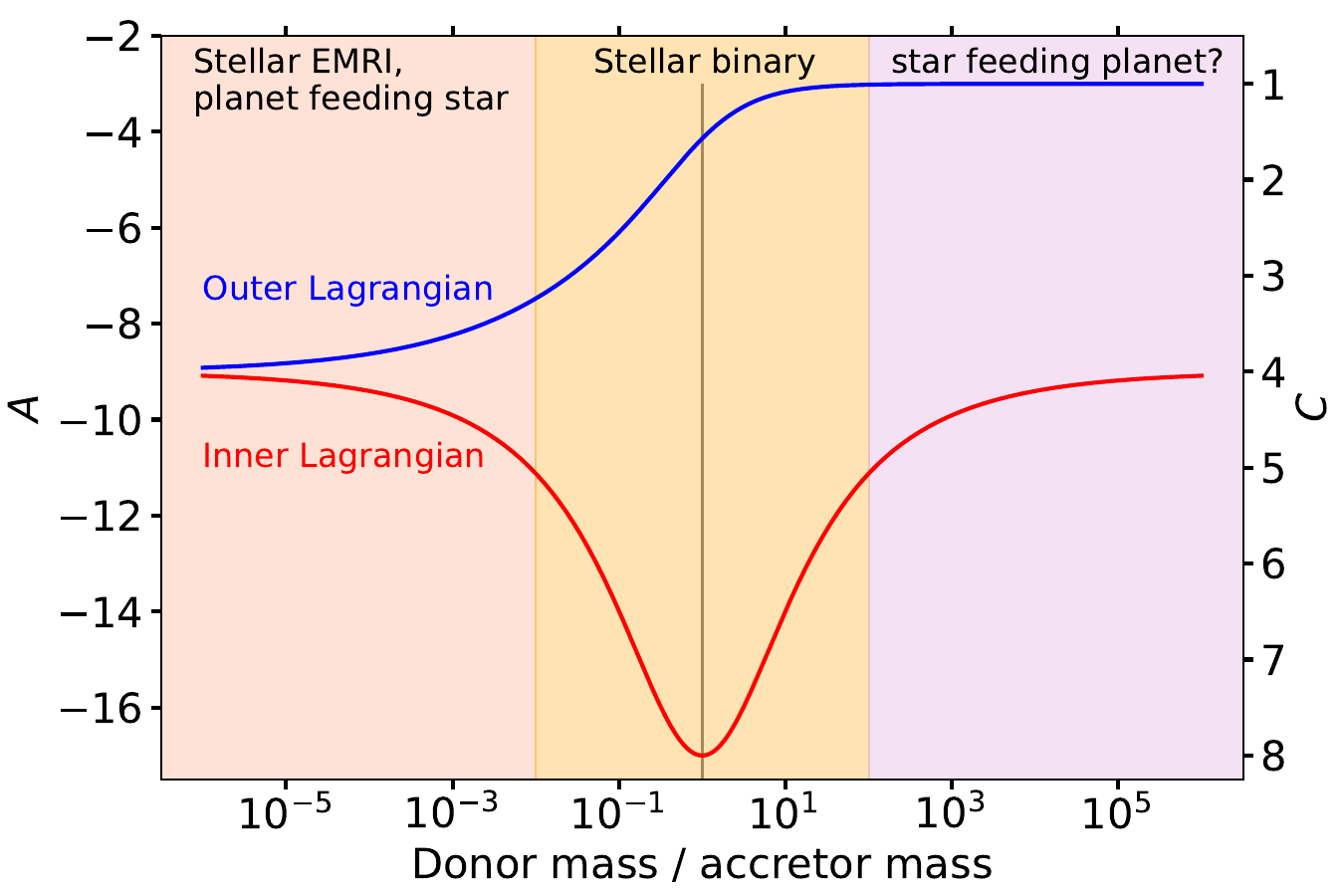}
    \includegraphics[width=0.48\linewidth]{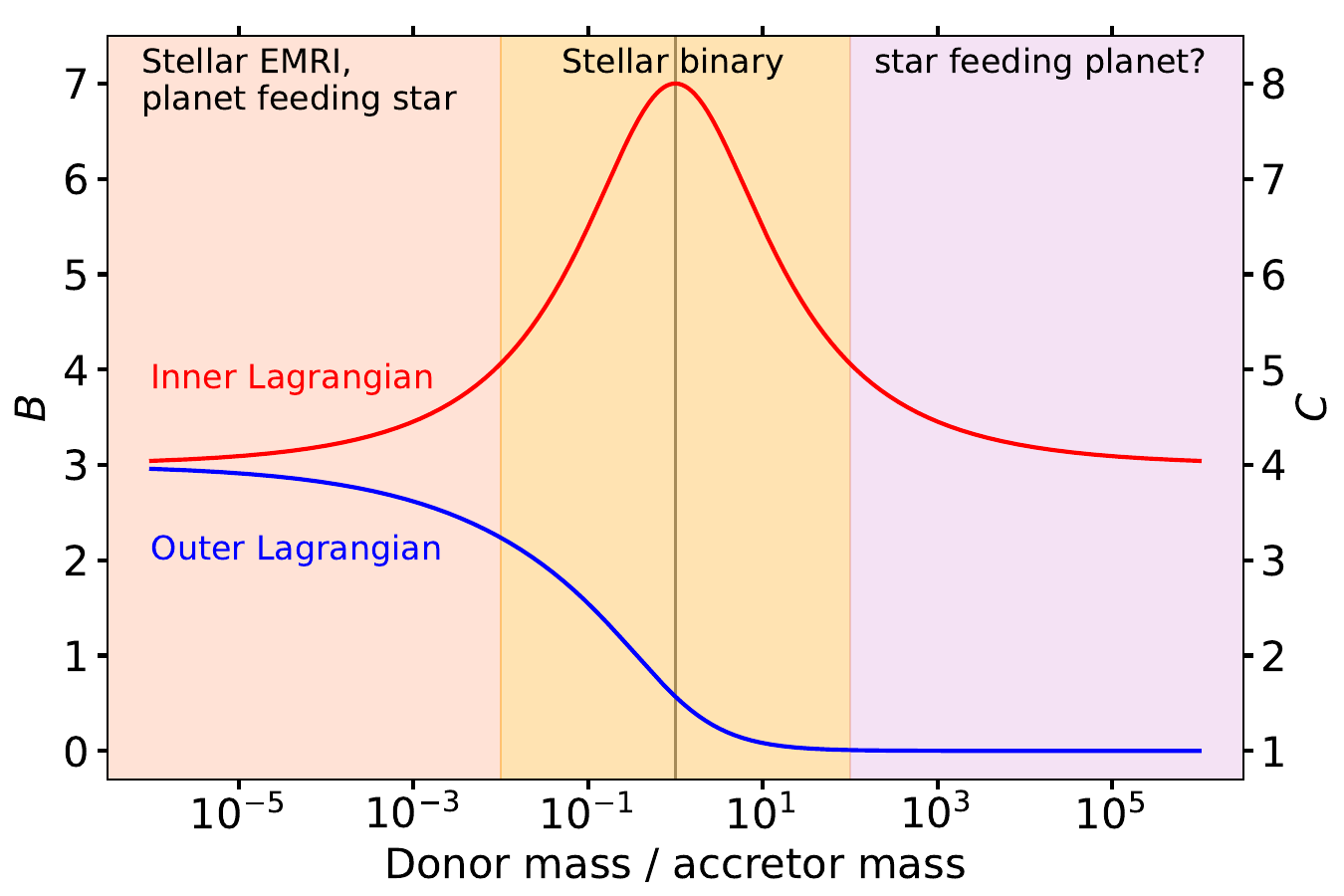}
    \caption{\label{fig:AB} Quadratic coefficients (Eq.~\ref{eq:phi}) that determine the curvature of the Roche potential near the Lagrangian point along the binary axis ($A$, left) and in the plane normal to the binary axis ($B$ and $C$, right). The three coefficients are related by $B = -(A+3)/2$ and $C=B+1$. The red (blue) line represents the coefficients for the inner (outer) Lagrangian point, as a function of the donor mass over the accretor mass. In both panels, a few examples of mass transferring systems are presented in three different mass ratio regimes such as stellar extreme mass ratio inspiral (EMRI), even though the boundaries among these regimes are not clearly defined.}
\end{figure*}

In this section, we derive analytic solutions for the internal structure of overcontact binaries in hydrostatic equilibrium with the Roche potential and a steady state overflow rate, assuming a polytropic relation. In addition, we introduce scaling factors to non-dimensionalize the problem. 

\subsection{Approximate Roche potential}\label{suubsec:RL}
In Sect.~\ref{sec:problem}, we introduced an approximate expression for the Roche potential $\phi$ (Eq.~\ref{eq:phi})
\begin{align}
    \phi(x,y,z) \simeq \frac{1}{2}\Omega^{2}(A x^{2} + B y^{2} + C z^{2}).
\end{align}
Note that while a higher-order expansion of the potential can be used, as done in \citet{Marchant+2021}, we adopt the second-order expansion for scalability.
The coefficients $A$, $B$, and $C$ can be computed numerically. In fact, in our hydrodynamical simulations, we used numerically computed values of those constants, which is shown in Fig.~\ref{fig:AB}. Notice that those coefficients for $L_{\rm in}$ are symmetric around $q=1$ (for example, $A$ and $B$ are the same for the mass ratios $q$ and $1/q$).  

While a fitting formula for $A$ was already provided in previous work \citep[for instance, Eq. 10 in ][]{Jackson+2017}, we also provide a fitting formula for $A$ sharing the same mathematical form of the fitting formulas for $\theta$, $\dot{\tilde{M}}/\dot{\tilde{M}}_{\rm analytic}$, and $\mathcal{F}_{1}$, accurate within $6\%$
\begin{align}\label{eq:fitting}
    A= a + b \tanh[c\log_{10}(q) + d]^{e},
\end{align}
where
\begin{align}\label{eq:A_fit}
    &a = -16.8, ~b = 7.53, ~c = 0.67, ~d=0, ~e=2  \hspace{0.1in} \text{for $L_{\rm in}$}\\
    &a = -5.87, ~b = -2.87, ~c = -0.68, ~d = -0.67, ~e=1 \hspace{0.1in}\text{for $L_{\rm out}$}.
\end{align}
Using Eq.~\ref{eq:A_fit}, $B=-(A+3)/2$ and $C=-(A+1)/2$ can be estimated. 

\subsection{Hydrostatic equilibrium solutions in the Roche potential}\label{subsec:analytic}
We first derive hydrostatic equilibrium solutions for the internal structure of a contact binary in the Roche potential $\phi$ (Eq.~\ref{eq:phi}), assuming a polytropic relation $P=K\rho^{\gamma}$ with a constant $K$ and a constant $\gamma$. The hydrostatic equilibrium condition can be written as
\begin{equation}\label{eq:hse}
    \grad \phi+\gamma K\rho^{\gamma-2}\grad \rho=0.
\end{equation}
Because the expression for $\rho$ is different, depending on whether $\gamma>1$ or $\gamma=1$, we derive the expression for each case separately. Accordingly, based on the functional form of each solution, we introduce different scaling factors to non-dimensionlize both the equilibrium solution and $\phi$. While the scaling factors are different, the expression for the scaled $\phi$ remains the same in both cases as
\begin{align}\label{eq:phiscaled}
    \tilde{\phi}(\tilde{x},\tilde{y},\tilde{z})=\frac{1}{2}(A \tilde{x}^{2} + B \tilde{y}^{2} + C \tilde{z}^{2}).
\end{align}
The tilde symbol, such as $\tilde{\phi}$, indicates dimensionless quantities in this paper.

\subsubsection{Adiabatic case with $\gamma>1$}
This case corresponds to overflowing streams that are optically thick (for instance, photosphere outside RL). 
Eq.~\ref{eq:hse} can be rewritten as
\begin{align}
\label{eq:hsegl1}
   \grad(\phi+{\gamma\over \gamma-1} K \rho^{\gamma-1})=0,
\end{align}
or
\begin{equation}\label{eq:hse2}
    \phi+{\gamma\over \gamma-1} K \rho^{\gamma-1}=\zeta,
\end{equation}
where $\zeta$ is a constant with the same dimension as $\phi$. $\zeta$ corresponds to the potential at the surface of the donor ($\phi$ at $\rho = 0$), which is equivalent to the overfilling factor of the donor star. Using Eq.~\ref{eq:phi}, and scaling the distance by $\Omega^{-1}\sqrt{|\zeta|}$ and the density by $(\gamma K/ (\gamma-1)|\zeta|)^{1/(\gamma-1)}$, we obtain the scaled expression for Eq.~\ref{eq:hse2}
\begin{align}
\label{eq:desnsitygl1}
    \tilde{\phi} + \tilde{\rho}^{\gamma-1}=\xi,
\end{align}
where, $\xi$ is either 1, corresponding to the overfilling case, or -1, corresponding to the underfilling case. Because we are interested in the overfilling case, we set $\xi=1$ from now on. Using Eq.~\ref{eq:phiscaled}, we find the expression for $\tilde{\rho}$
\begin{align}
    \label{eq:adia_rho}
    \tilde{\rho}=\left[ 1-{A \over 2} \tilde{x}^2-{B \over 2} \tilde{y}^2 -{C\over 2} \tilde{z}^2\right]^{1 \over \gamma-1}.
\end{align}

\begin{figure}
    \centering
    \includegraphics[width=9cm]{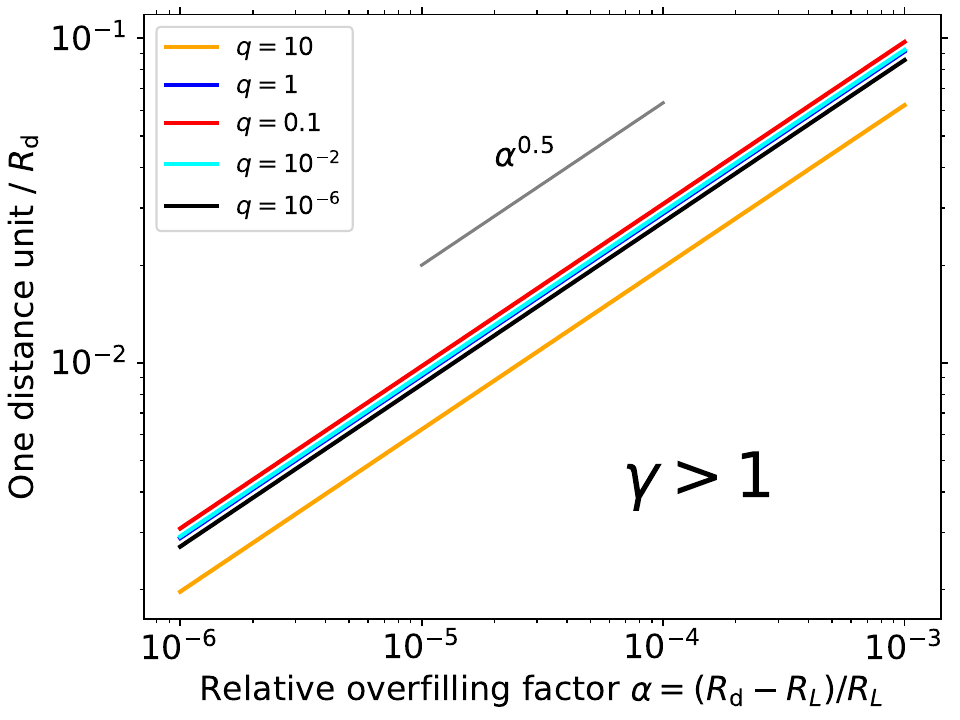}
    \caption{\label{fig:codeunit_Rd} Ratio of one distance code unit ($\Omega^{-1}\sqrt{|\xi|}$) to the effective size of the donor $R_{\rm d}$, as a function of the relative overfilling factor $\alpha$, defined as $\alpha=(R_{\rm d}-R_{\rm L})/R_{\rm L}$. Here, $R_{\rm L}$ is the effective RL size. We consider the inner Lagrangian point for this conversion.}
\end{figure}
We have formulated the problem such that the potential at the surface of the overfilling donor ($\xi$), or equivalently the overfilling factor, acts as a scaling factor. As mentioned previously, this formulation allows us to scale our results to an arbitrarily small overfilling factor. However, the relationships between the overfilling factor and other scaling factors become less obvious in the scaled equations when interpreting our hydrodynamics simulation results in code units (see Sect.~\ref{sec:methods}). As an example, we illustrate how the relative overfilling factor $\alpha$ scales with the distance scaling factor for $L_{\rm in}$ in Fig.~\ref{fig:codeunit_Rd}. For 
 a given $\alpha$, we estimate $\phi$ at the donor surface relative to $\phi$ at the Lagrangian point using the higher-order expansion of the equipotential derived by \citet{Jackson+2017}. We also estimate $R_{\rm d}$ using the same expression for the potential. 
The distance scaling factor, in units of $R_{\rm d}$, depends weakly on $\alpha$, namely, $\simeq (0.6-1)\times10^{-2}(\alpha/10^{-5})^{0.5}$ for $L_{\rm in}$ within $10^{-6}\lesssim q\lesssim 10$.

\subsubsection{Isothermal case with $\gamma=1$}
The isothermal case corresponds to the scenario where the transferring mass is optically thin (for instance, photosphere inside RL), maintaining the same sound speed. For this case, the equivalent form of Eq.~\ref{eq:hsegl1} is
\begin{equation}
    \grad (\phi+K \ln \rho)=0,
\end{equation}
which yields
\begin{equation}
    \rho = \rho_0 \exp \left( -\phi /K \right).
\end{equation}
Scaling the distance by $\sqrt{K}\Omega^{-1}$ and the density by the arbitrary factor $\rho_0$, we obtain
\begin{equation}\label{eq:iso_rho}
    \tilde {\rho}=\exp \left[ -{A \over 2} \tilde x^2 - {B \over 2} \tilde{y}^2 -{C \over 2} \tilde{z}^2  \right].
\end{equation}
Because of the infinite extent of the envelope, such donors  always overfill the RL.

\subsection{Stationary solutions for mass overflow rate}\label{subsec:ana_mdot}
In this section, we follow the assumptions made in \citet{Ritter1988} for the isothermal case and \citet{KolbRitter1990} for the adiabatic case to derive the stationary solutions for the mass overflow rate. In the following derivation, we do not include the Coriolis force.

\subsubsection{Adiabatic case with $\gamma>1$}
We start by considering the mass transfer rate exactly at the Lagrangian point. The corresponding 1D continuity equation integrates to 
\begin{equation}\label{eq:1Dcontinuity}
\rho v = \dot{M}_{\rm L} = {\rm constant}.
 \end{equation}
Notice that $\dot{M}_{\rm L}$ in Eq.~\ref{eq:1Dcontinuity} has the same units as mass flux [mass / time / area ], not [mass / time]. However, for notational consistency, we continue to use the symbol $\dot{M}_{\rm L}$. 
Using Eq.~\ref{eq:1Dcontinuity} and assuming the polytropic relation, the momentum equation can be rewritten as
\begin{align}
\frac{D}{Dt}\left(\frac{\dot{M}_{\rm L}}{\rho}\right) & =-\nabla\left(\phi+\frac{\gamma}{\gamma-1}K\rho^{\gamma-1}\right),\\
0 & =-\nabla\left(\phi+\frac{\gamma}{\gamma-1}K\rho^{\gamma-1}+\frac{\dot{M}_{\rm L}^{2}}{2}\rho^{-2}\right),
\end{align}
where $D/Dt=\partial/\partial t+v\partial/\partial x$. As this is a full derivative we obtain
\begin{equation}\label{eq:Ber1D}
    \phi+\frac{\gamma}{\gamma-1}K\rho^{\gamma-1}+\frac{\dot{M}_{\rm L}^{2}}{2}\rho^{-2}=\phi_0={\rm constant}.
\end{equation}
This is known as the Bernoulli equation. Notice that $\phi_{0}=\zeta$ when $\dot{M}=0$ (hydrostatic equilibrium, Eq.~\ref{eq:hse2}). At the given Lagrangian point (either $L_{\rm in}$ or $L_{\rm out}$), the condition $\nabla\phi \notag=0$ gives
\begin{equation}\label{eq:mdotL}
\dot{M}_{\rm L}^{2}=(\rho_{\rm L}c_{\rm s,L})^{2}= \gamma K\rho_{\rm L}^{\gamma+1},
\end{equation}
where $\rho_{\rm L}$ is the density and $c_{\rm s,L}$ is the sound speed, defined as $\sqrt{\gamma P_{\rm L}/\rho_{\rm L}}$, at the Lagrangian point. Note that $\nabla\phi \notag=0$ indicates that Eq.~\ref{eq:mdotL} expresses the maximum $\dot{M}_{\rm L}$, where the transferring stream speed equal the sound speed, or $v_{\rm L}=c_{\rm s,L}$. This further implies that if $v_{\rm L}\neq c_{\rm s,L}$ (either sub or supersonic), the resulting flux will be smaller than that estimated using Eq.~\ref{eq:mdotL}.

Inserting the expression for $\dot{M}_{\rm 1D}$ into Eq.~\ref{eq:Ber1D}, we finally obtain
\begin{align}\label{eq:1DrhoL}
    \rho_{\rm L}=\left[ \frac{ 2(\gamma-1) }{\gamma(\gamma+1)} \frac{\phi_0-\phi_{\rm L}}{K} \right]^{\frac{1}{\gamma-1}},
\end{align}
\begin{align}\label{eq:1DvxL}
    v^{x}_{\rm L}= \sqrt{\frac{2(\gamma-1)}{\gamma+1}(\phi_0-\phi_{\rm L})},
\end{align}
and
\begin{equation}
\label{eq:Mdot1D}
\dot{M}_{\rm L}=(\phi_{0}-\phi_{\rm L})^{\frac{\gamma+1}{2\left(\gamma-1\right)}}{\left(\gamma K\right)}^{-\frac{1}{\gamma-1}}\left[\frac{2(\gamma-1)}{\gamma+1}\right]^{\frac{\gamma+1}{2\left(\gamma-1\right)}}.
\end{equation}
It is worth pointing out that the mass flux is independent of the shape of the potential $\phi$, but depends only on the difference in the potential $\phi_0-\phi_{\rm L}$. 
This can be compared with an order-of-magnitude estimate for the mass flux $\dot{M}_{\rm L, approx}$ as the product of the hydrostatic solutions for the density and the sound speed at the Lagrangian point (using Eq.~\ref{eq:hse2} rather than Eq.\ref{eq:Ber1D}), resulting in the ratio of $\dot{M}_{\rm L}/\dot{M}_{\rm L, approx}$ as
\begin{equation}
\label{eq:1Derror}
\frac{\dot{M}_{\rm L}}{\dot{M}_{\rm L,approx}}=\left(\frac{2}{\gamma+1}\right)^{\frac{\gamma+1}{2\left(\gamma-1\right)}}.
\end{equation}
For $\gamma=5/3$, the hydrostatic equilibrium condition overestimates the overflowing rate by a factor of 1.78.

Now the mass overflow rate across the plane, normal to the binary axis, intersecting the Lagrangian point can be expressed as
\begin{equation}
    \dot{M}_{\rm L}=\int\int\rho(y,z)_{\rm L} v^{\rm x}(y,z)_{\rm L} {\rm d}y {\rm d}z.
\end{equation}
If the Bernoulli constant is the same across the overflowing stream, we can still use the same expression for $\rho_{\rm L}$ and $v^{\rm x}_{\rm L}$ as Eqs.~\ref{eq:1DrhoL} and~\ref{eq:1DvxL}, respectively, with $\phi_{\rm L}$ dependent on $y$ and $z$. The mass overflow rate is
\begin{align}
    \dot{M}_{\rm L}&=4\pi\left(\frac{\gamma-1}{3\gamma-1}\right)\left(\gamma K\right)^{-\frac{1}{\gamma-1}}\cdot\left(\frac{2\left(\gamma-1\right)}{\gamma+1}\right)^{\frac{\gamma+1}{2\left(\gamma-1\right)}}\frac{(\phi_{0}-\phi_{\mathrm{L}})^{\frac{3\gamma-1}{2\left(\gamma-1\right)}}}{\sqrt{BC}}.
\end{align}
The scaled overflow rate is
\begin{align}\label{eq:mdot_3Dadiabatic}
    \dot{\tilde{M}}_{\rm L} = \frac{ 4\pi}{\sqrt{BC}}\left[\frac{(\gamma-1)^{3/2}}{3\gamma-1}\right]\left[\frac{2}{\gamma+1}\right]^{\frac{\gamma+1}{2\left(\gamma-1\right)}}.
\end{align}
\subsubsection{Isothermal case with $\gamma=1$}
We derive the solution for the isothermal case in a similar way for the adiabatic case with $\gamma>1$. The 1D Bernoulli equation for $\gamma=1$ is
\begin{equation}\label{eq:Ber1Diso}
    \phi+ K\ln\rho+\frac{\dot{M}_{\rm 1D}^{2}}{2}\rho^{-2}=\phi_0={\rm constant},
\end{equation}
which gives
\begin{align}\label{eq:mdotL_iso}
    \dot{M}^{2}_{\rm L}=K\rho_{\rm L}^{2},
\end{align}
and 
\begin{align}
    \rho_{\rm L} =\rho_0 \exp[\frac{\phi_{0}-\phi_{\rm L}-K/2}{K}].
\end{align}
Similarly to the adiabatic case, Eq.~\ref{eq:mdotL_iso} expresses the maximum rate, where the Mach number of the transferring stream is unity at the Lagrangian point. Extending this solution to that for the stream through the 2D plane intersecting the Lagrangian point in a manner similar to what we have done for the adiabatic case, we find
\begin{align}
    \dot{M}_{\rm L}=\frac{2\pi\rho_0 K^{3/2}}{\Omega^{2}\sqrt{BC}}\exp[\phi_{0}/K-1/2],
\end{align}
and the scaled rate is
\begin{align}\label{eq:mdot_3Disothermal}
    \dot{\tilde{M}}_{\rm L}=\frac{2\pi\sqrt{e}}{\sqrt{BC}}.
\end{align}
While \citet{Ritter1988} introduced a fitting formula ($F(q)$ in their Eq.~A9) for estimating $\sqrt{BC}$, we compute the term numerically.

\subsection{Scaling factors}\label{subsec:scalingfactors}

We summarize the scaling factors introduced above:
\begin{itemize}
\item The adiabatic case $P=K\rho^{\gamma}$ with $\gamma>1$: $\sqrt{\zeta}$ for velocity, $\Omega^{-1}$ for time, $\Omega^{-1}\sqrt{|\zeta|}$ for distance, and $\left[\frac{(\gamma -1)|\zeta|}{\gamma K}\right]^{1/(\gamma-1)}$ for density. \\

\item The isothermal case $P=K\rho$: $\sqrt{K}$ for velocity, $\Omega^{-1}$ for time, $\Omega^{-1}\sqrt{K}$ for distance, and $\rho_{0}$ for density. 
\end{itemize}

Using these scaling relations, scaled quantities can be converted into  physical units.

\section{Numerical methods}\label{sec:methods}
\begin{figure}
    \centering
    \includegraphics[width=8.8cm]{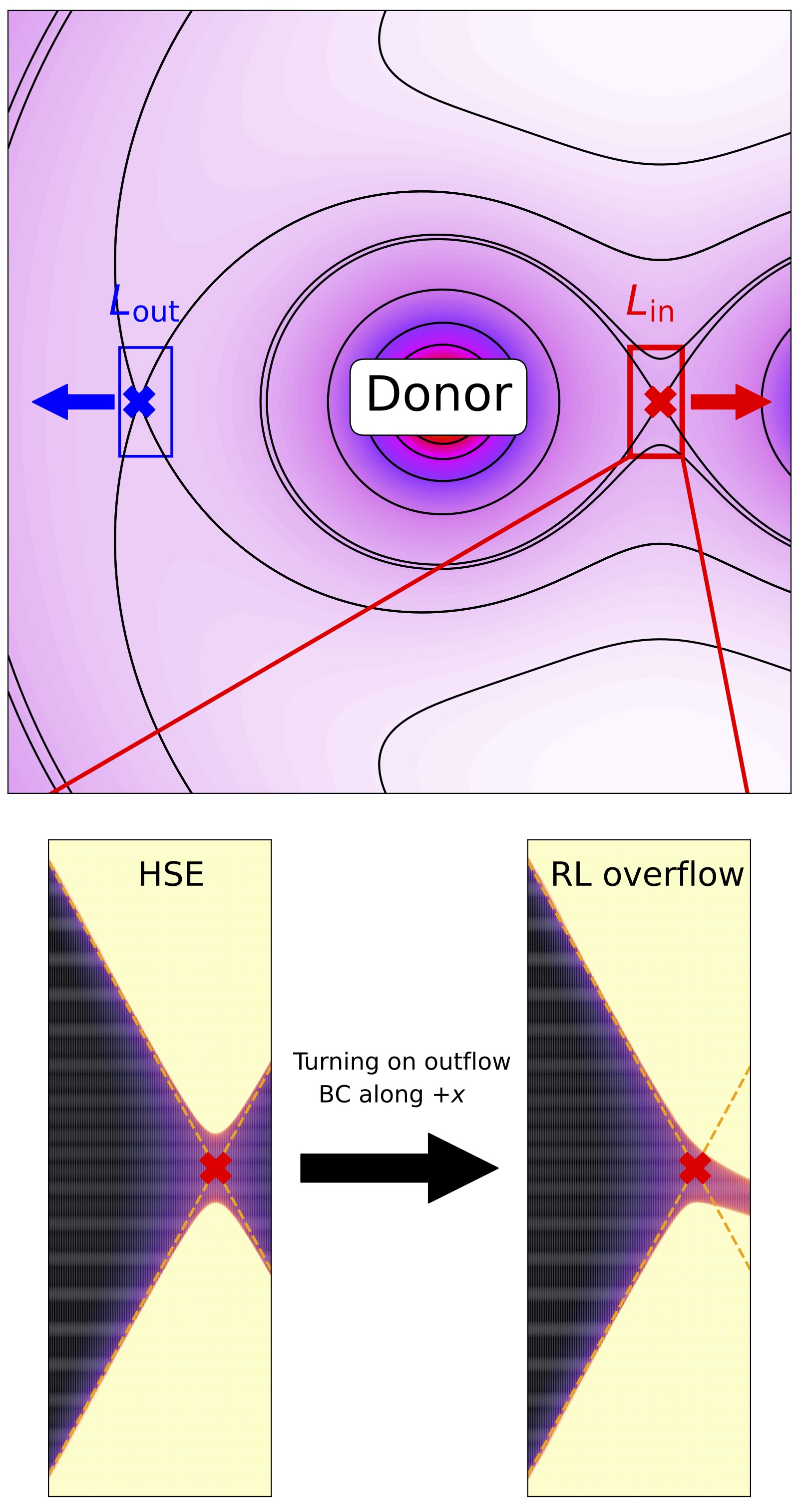}
    \caption{\label{fig:schematic} Schematic diagram for the location of our computational domain and transition from a hydrostatic solution to Roche lob overflow. The top panel depicts the shape of the Roche-lobe equipotential surfaces, along with the approximate locations of our computational domains (rectangles) around $L_{\rm in}$ (red cross) and $L_{\rm out}$ (blue cross). The arrows indicate the direction of overflowing gas. The diagram is drawn not to scale. The bottom panels depict the density change, with denser regions shown in darker colors, during the transition from a contact binary in hydrostatic equilibrium to overflow by imposing an outflow boundary condition (BC) along the right $x$ boundary. The dashed diagonal orange lines depict the shape of the RL. }
\end{figure}

\subsection{Code setup}\label{appendix:codesetup}
To find numerical solutions for the mass overflow, we solved 3D hydrodynamics equations with the scaled Roche potential (Eq.~\ref{eq:phiscaled}) and the Coriolis force near the Lagrangian point in the corotating frame, using the finite-volume adaptive mesh refinement magnetohydrodynamics code {\small ATHENA++} \citep{Stone+2008,Stone+2020}. In this work, we did not include magnetic fields. We adopted the 3rd order spatial reconstruction scheme -- the highest order supported with adaptive mesh refinement -- with the 3rd-order Runge-Kutta time integration algorithm. 

The computational domain is a rectangular cuboid, with extents along the $y$ and $z$ axes longer than that along the $x$ axis. Fig.~\ref{fig:schematic} depicts the location and shape of the domain around the $L_{\rm in}$ and $L_{\rm out}$ points. While the Lagrangian point is always located at the coordinate origin, it is not always positioned at the geometric center of the domain. The $x$-extent toward the donor's center ($\tilde{x}_{\rm min}$) determines the validity of our approximate potential and the density range covered in the domain. For $\gamma=5/3$, the left boundary is located at $\tilde{x}=-3$ and the right boundary at $\tilde{x}=1$, independent of $q$. For this size, the departure of the approximate potential from the exact value generally $\lesssim 1\%$ for an overfilling factor $\lesssim 10^{-2}$. For $\gamma=1$, the density increases much more sharply, causing the density range to become too wide across the domain. For example, at $\tilde{x}=-3$, $\tilde{\rho}_{\tilde{x}=-3}/\tilde{\rho}_{\tilde{x}=0}\simeq 10^{33}$. Such a wide density range naturally leads to higher computational costs. So, we determined $\tilde{x}_{\rm min}$ such that the maximum density is $10^{3}-10^{4}$. For both adiabatic and isothermal cases, we extended the domain along $\tilde{y}$ and $\tilde{z}$ to fully enclose the gas at $\tilde{x}=\tilde{x}_{\rm min}$. 

To resolve the overflowing stream and stably maintain its steady state deep inside the donor, we adaptively refined the cells with $\tilde{\rho}>10^{-3}$. The base-level resolution and refinement levels (up to 2 levels for $\gamma=1$ and 3 levels for $\gamma=5/3$) were determined to reach the convergent gas morphology and mass overflow rate. At the base-level, we placed $10-20$ cells across the width of the stream crossing over the Lagrangian point, which increased by a factor of $2 - 4$ at the highest level of refinement in each direction. This resolution at the highest level of refinement corresponds to at least $30-50$ cells and $10-30$ cells per pressure scale height ($P/(dP/dR)$) inside the donor for the adiabatic and isothermal cases, respectively. We performed resolution tests for the equal-mass, $L_{\rm in}$ adiabatic and isothermal cases by varying the resolution at the highest level of refinement, and we confirmed that the resolution used in the simulations presented in the paper yield converged results. 

We adopted a simple $\Gamma$ equation of state, $P = (\Gamma-1)u$ with $\Gamma=5/3$ for $\gamma = 5/3$, and an isothermal equation of state for $\gamma=1$.

\begin{table}
\caption{Model parameters of our hydrodynamics simulations for RL overflow.}\label{tabl:models}
\centering
\small
\begin{tabular}{ c | c |c c c c} 
\hline
\hline
$\gamma$  & $L$ & $q$  & $N_{\rm base}$ & $\tilde{x}$ & $\tilde{y}$ \& $\tilde{z}$\\
\hline
 \multirow{14}{*}{$5/3$}&  \multirow{6}{*}{$L_{\rm in}$}  & $1^{\star}$ &  (80, 160, 160) & (-3,1) & (-5.0,5.0) \\
                        &                         & $0.3$ &  (80, 160, 160) & (-3,1) & (-5.0,5.0)\\
                        &                         & $0.1$ &  (72, 144, 144) & (-3,1) & (-5.1,5.1) \\
                        &                         & $10^{-2}$ &  (64, 128, 128) & (-3,1) & (-5.3,5.3)\\
                        &                         & $10^{-4}$ &  (64, 128, 128) & (-3,1) & (-5.5,5.5) \\
                        &                         & $10^{-6}$ &  (64, 128, 128) & (-3,1) & (-5.6,5.6)\\\cline{2-6}
                        &  \multirow{8}{*}{$L_{\rm out}$}   & $10$ &   (96, 192, 192) & (-3,1) & (-20.4,20.4) \\
                        &                          & $3.33$ &  (40, 160, 160) & (-3,1) & (-12.9,12.9) \\
                        &                          & $1  $ &  (64, 128, 128) & (-3,1) & (-8.8,8.8) \\
                        &                          & $0.3  $ &  (64, 128, 128) & (-3,1) & (-7.1,7.1) \\
                        &                          & $0.1 $ &  (64, 128, 128) & (-3,1) & (-6.4,6.4) \\
                        &                          & $10^{-2} $ &  (64, 128, 128) & (-3,1) & (-5.9,5.9) \\
                        &                          & $10^{-4} $ &  (64, 128, 128) & (-3,1) & (-5.7,5.7) \\
                        &                          & $10^{-6} $ &  (64, 128, 128) & (-3,1) & (-5.6,5.6) \\
\hline
\multirow{14}{*}{$1$}&  \multirow{6}{*}{$L_{\rm in}$}  & $1$ &  (64, 128, 128) & (-1.1,0.5) & (-2.6,2.6) \\
                        &                         & $0.3$ &  (72, 144, 144) & (-1.0,0.5) & (-2.7,2.7)\\
                        &                         & $0.1$ &  (64,128,128) & (-1.1,0.5) & (-2.9,2.9) \\
                        &                         & $10^{-2}$ &  (64,128,144) & (-1.2,0.5) & (-3.4,3.4)\\
                        &                         & $10^{-4}$ &  (64, 128, 128) & (-1.3,0.5) & (-3.8,3.8) \\
                        &                         & $10^{-6}$ &  (56, 112, 112) & (-1.3,0.5) & (-3.9,3.9)\\\cline{2-6}
                        &  \multirow{8}{*}{$L_{\rm out}$}   & $10$ &   (160, 352, 352) & (-2.1,2.0) & (-23.1,23.1) \\
                        &                          & $3.33$ &  (160, 336, 336) & (-2.1,2.0) & (-14.2,14.2) \\
                        &                          & $1  $ &  (128, 256, 256) & (-1.9,0.5) & (-8.9,8.9) \\
                        &                          & $0.3  $ &  (40, 240, 240) & (-1.7,0.5) & (-6.4,6.4) \\
                        &                          & $0.1 $ &  (48, 192, 192) & (-1.6,0.5) & (-5.4,5.4) \\
                        &                          & $10^{-2} $ &  (48, 192, 192) & (-1.4,0.5) & (-4.5,4.5) \\
                        &                          & $10^{-4} $ &  (72,144,144) & (-1.3,0.5) & (-4.0,4.0) \\
                        &                          & $10^{-6} $ &  (72,144,144 ) & (-1.3,0.5) & (-3.9,3.0) \\
 \hline
\end{tabular}
\tablefoot{From left to right, the column gives the polytropic exponent $\gamma$, the location of the Lagrangian point, the mass ratio $q$, the number of cells at the base-level, the domain extent along the $x$ axis, and the domain extent along the $y$ and $z$ axes. The symbol $^{\star}$ indicates the model for which we perform the simulation both with and without the Coriolis force. }
\end{table}
\subsection{From a contact binary to RL overflow -- Mapping, relaxation, and boundary conditions}

We first mapped the solution for a contact binary onto a 3D domain using Eq.~\ref{eq:adia_rho} for $\gamma=5/3$ or Eq.~\ref{eq:iso_rho} for $\gamma=1$, and then relaxed it without the Coriolis force. The solution for $\gamma=5/3$ has a sharp edge at which $\tilde{\rho}=0$. For numerical stability, we impose a density floor of $\tilde{\rho}=10^{-7}$, effectively extending the edge only down to this floor value. In addition, the rest of the domain is filled with low-density gas at the floor value, with the pressure determined by the polytropic relation. The solution for $\gamma=1$ does not have a sharp edge. But we still impose a density floor of $10^{-10}$. 

To efficiently damp the motion of gas during the relaxation phase and in the transition phase once the gas begins to flow, we employed an isotropic viscosity, which depends on time and location, such that
\begin{align}
    \nu(\tilde{t}, \tilde{r}) = \frac{\tanh[-3(sign(\tilde{x})\tilde{r}+1)] + 1 }{2} \left[\nu_{0}(\tanh(-\tilde{t}^{2}/t_{\nu}) + 1) + \nu_{\rm min}\right],
\end{align}
where $\tilde{r}$ is the radial distance from the Lagrangian point. 
The first term determines the location dependence of $\nu$. The first term is of order unity in the inside of the donor ($\tilde{r}>1$ and $sign(\tilde{x})=-1$), which approaches zero toward the accretor ($sign(\tilde{x})=1$). The second term considers the temporal evolution of $\nu$ such that the viscosity deceases from $\nu_{0}$ to $\nu_{\rm min}$ over a timescale of $t_{\rm nu}$. To maintain reasonable step-sizes, we employed super-time-stepping method with the second order Runge-Kutta-Legendre integrator. We took $\nu_{0}\simeq 0.1-0.5$, $t_{\nu}=3-4$, and $v_{\rm min}=10^{-2}-10^{-3}$, depending on $q$ and whether it is for the $L_{\rm in}$ or $L_{\rm out}$ point. These values for viscosity, which were chosen based on comprehensive test simulations, are initially high enough to efficiently damp the transient motion of gas in a transition phase and then quickly drops so that the overflowing rate reaches a steady state. If the viscosity remains high for too long, the overflow rate tends to continuously decrease, rather than reaching a steady state solution.

Under these conditions, relaxation usually took around $\tilde{t}\simeq 2 - 3$ to damp numerical oscillations of gas within the domain. The Mach number of fully relaxed gas was $\lesssim 10^{-2}$ in the domain. The boundary conditions at this stage were given such that the ghost cells were filled, following the analytic solutions (Eq.~\ref{eq:adia_rho} or \ref{eq:iso_rho}) if the ``real'' gas initially touched the boundary, and with zero-gradient extrapolation for the rest of the boundaries. 

Once the gas was fully relaxed, we applied the outflow boundary condition to the positive $x$-boundary (toward the accretor) and included the Coriolis force, while the density floor values and the rest of the boundary conditions remained the same. This effectively emulates the RL overflow in a semi-detached binary by driving gas outflows beyond the Lagrangian point from an overfilling donor through the positive $x$-boundary. The transition from a contact binary to RL overflow is shown in the bottom panels in Fig.~\ref{fig:schematic}. The transferring gas and the donor's surface typically underwent a transition phase before reaching a steady state typically within $\tilde{t}\simeq 5 - 15$, which is sufficiently shorter than the duration of simulations ($\tilde{t}\gtrsim 50$ for $q\leq 1$ and $25-30$ for $q>1$).

\end{appendix}

\end{document}